\documentclass[longauth]{aa} 

\usepackage{graphicx}
\usepackage{txfonts}
\usepackage{lineno} 
\usepackage[colorlinks=true,linkcolor=bluea,allcolors=blue]{hyperref}
\newcommand{\barolo}{\textsuperscript{3D}BAROLO}

\begin{document}

  \title{Molecular gas stratification and disturbed kinematics in the Seyfert galaxy MCG-05-23-16 revealed by JWST and ALMA}
   \author{D. Esparza-Arredondo
          \inst{1,2}
          \thanks{\email{donaji.esparza@iac.es}}
          \and
          C. Ramos Almeida
          \inst{1,2}
          \and
          A. Audibert
          \inst{1,2}
          \and
          M. Pereira-Santaella
          \inst{3}
          \and
          I. García-Bernete
          \inst{4}
          \and
         S. García-Burillo
        \inst{5}
          \and
          T. Shimizu
          \inst{6}
          \and
          R. Davies
        \inst{6}
        \and
        L. Hermosa Muñoz
        \inst{4}
        \and
        A. Alonso-Herrero
        \inst{4}
        \and
        F. Combes
        \inst{7}
        \and
        G. Speranza
        \inst{3}
        \and
        L. Zhang
        \inst{8}
        \and
        S. Campbell
        \inst{9}
        \and
        E. Bellocchi
        \inst{10,11}
        \and
        A. J. Bunker
        \inst{12}
         \and
        T. Díaz-Santos
        \inst{13,14}
        \and
        B. García-Lorenzo
        \inst{1,2}
        \and
        O. González-Martín
        \inst{15} 
         \and
        E. K. S. Hicks
        \inst{8,16,17}
        \and
        A. Labiano
        \inst{18}
        \and
        N.A. Levenson
        \inst{19}
        \and
        C. Ricci
        \inst{20,21}
        \and
        D. Rosario
        \inst{9}
        \and
        S. Hoenig
        \inst{22}
        \and
        C. Packham
        \inst{8}
        \and 
        M. Stalevski
        \inst{23,24}
        \and
        L. Fuller
        \inst{9}
        \and
        T. Izumi
        \inst{25}
        \and
        E. López-Rodríguez
        \inst{26,27}
        \and
        D. Rigopoulou
        \inst{4,14}
        \and
        D. Rouan
        \inst{27}
        \and
        M. Ward
        \inst{28}
          }
        \institute{Instituto de Astrofísica de Canarias, Calle Vía Láctea, s/n, E-38205, La Laguna, Tenerife, Spain
         \and
             Departamento de Astrofísica, Universidad de La Laguna, E-38206, La Laguna, Tenerife, Spain
          \and
            Instituto de Física Fundamental, CSIC, Calle Serrano 123, 28006
Madrid, Spain
           \and
            Centro de Astrobiología (CAB), CSIC-INTA, Camino Bajo del
Castillo s/n, 28692 Villanueva de la Cañada, Madrid, Spain
          \and
           Observatorio Astronómico Nacional (OAN-IGN)-Observatorio de
Madrid, Alfonso XII, 3, 28014 Madrid, Spain
            \and
            Max Planck Institute for Extraterrestrial Physics (MPE), Giessenbachstr.1, 85748 Garching, Germany
            \and
            LERMA, Observatoire de Paris, Collège de France, PSL University, 837 CNRS, Sorbonne University, Paris
            \and
            Department of Physics and Astronomy, The University of Texas at San Antonio, 1 UTSA Circle, San Antonio, Texas, 78249, USA
           \and
           School of Mathematics, Statistics and Physics, Newcastle University, Newcastle upon Tyne, NE1 7RU, UK
            \and
           Departmento de Física de la Tierra y Astrofísica, Fac. de CC Físicas, Universidad Complutense de Madrid, E-28040 Madrid, Spain
           \and
           Instituto de Física de Partículas y del Cosmos IPARCOS, Fac. CC Físicas, Universidad Complutense de Madrid, E-28040 Madrid, Spain
           \and
            Department of Physics, University of Oxford, Keble Road, Oxford OX1 3RH, UK
           \and
            Institute of Astrophysics, Foundation for Research and Technology Hellas (FORTH), Heraklion 70013, Greece
            \and
            School of Sciences, European University Cyprus, Diogenes street, Engomi, 1516 Nicosia, Cyprus
           \and
            Instituto de Radioastronomía and Astrofísica (IRyA-UNAM), 3-72 (Xangari), 8701 Morelia, Mexico
            \and
            Department of Physics \& Astronomy, University of Alaska Anchorage, Anchorage, Alaska, USA
            \and
            Department of Physics, University of Alaska, Fairbanks, Alaska 99775-5920, USA
            \and
            Telespazio UK for the European Space Agency (ESA), ESAC, Camino Bajo del Castillo s/n, 28692 Villanueva de la Cañada, Spain
            \and
            Space Telescope Science Institute, 3700 San Martin Drive, Baltimore, MD 21218, USA
            \and
            Instituto de Estudios Astrofísicos, Facultad de Ingeniería y Ciencias, Universidad Diego Portales, Av. Ejército Libertador 441, Santiago, Chile
            \and
            Kavli Institute for Astronomy and Astrophysics, Peking University, Beijing 100871, China
            \and
            School of Physics \& Astronomy, University of Southampton, High- 843 field, Southampton SO17 1BJ, UK
            \and
            Astronomical Observatory, Volgina 7, 11060 Belgrade, Serbia.
            \and
            Sterrenkundig Observatorium, Universiteit Gent, Krijgslaan 281-S9, Gent B-9000, Belgium
            \and
            Department of Astronomy, School of Science, Graduate University for Advanced Studies (SOKENDAI), Mitaka, Tokyo 181-8588, Japan
            \and
            Kavli Institute for Particle Astrophysics \& Cosmology (KIPAC), Stanford University, Stanford, CA 94305, USA
            \and
            Department of Physics \& Astronomy, University of South Carolina, Columbia, SC 29208, USA
            \and
            LESIA, Observatoire de Paris, Universite PSL, CNRS, Sorbonne Universite, Sorbonne Paris Cite'e, 5 place Jules Janssen, F-92195 Meudon, France
            \and
            Centre for Extragalactic Astronomy, Durham University, South Road, Durham DH1 3LE, UK
            }
    
   \date{Received October 4, 2024; Accepted November 18, 2024; }

   \titlerunning{Molecular gas stratification and disturbed kinematics in MCG-05-23-16}

 
  \abstract
   {Understanding the processes that drive the morphology and kinematics of molecular gas in galaxies is crucial for comprehending star formation and, ultimately, galaxy evolution. Using data obtained with the James Webb Space Telescope (JWST) and the Atacama Large Millimeter/submillimeter Array (ALMA), we study the behavior of the warm molecular gas at temperatures of hundreds of Kelvin and the cold molecular gas at tens of Kelvin in the galaxy MCG$-$05$-$23$-$16, which hosts an active galactic nucleus (AGN). Hubble Space Telescope (HST) images of this spheroidal galaxy, classified in the optical as S0, show a dust lane resembling a nuclear spiral and a surrounding ring. These features are also detected in CO(2$-$1) and H$_2$, and their morphologies and kinematics are consistent with rotation plus local inward gas motions along the kinematic minor axis in the presence of a nuclear bar. The H$_2$ transitions 0-0 S(3), 0-0 S(4), and 0-0 S(5), which trace warmer and more excited gas, show more disrupted kinematics than 0-0 S(1) and 0-0 S(2), including clumps of high-velocity dispersion (of up to $\sim$160 km~s$^{-1}$), in regions devoid of CO(2$-$1). The kinematics of one of these clumps, located at $\sim$350\,pc westward from the nucleus, are consistent with outflowing gas, possibly driven by localized star formation traced by Polycyclic Aromatic Hydrocarbon (PAH) emission at 11.3 $\mu$m. Overall, we observe a stratification of the molecular gas, with the colder gas located in the nuclear spiral, ring, and connecting arms, while most warmer gas with higher velocity-dispersion fills the inter-arm space. The compact jet, approximately 200 pc in size, detected with Very Large Array (VLA) observations, does not appear to significantly affect the distribution and kinematics of the molecular gas, possibly due to its limited intersection with the molecular gas disc.}
   \keywords{galaxies: active -- galaxies: nuclei -- galaxies: individual:MCG-05-23-16 -- galaxies: Seyfert -- galaxies:ISM -- galaxies:kinematics and dynamics}
   \maketitle

\section{Introduction}
\label{sec:Introduction}

A wealth of observational evidence supports the co-evolution of Active Galactic Nuclei (AGN) and their host galaxies. This includes the various correlations between the supermassive black hole (SMBH) mass (M$_{\rm BH})$ and different galaxy properties, such as the velocity dispersion and the mass of the bulge \citep[e.g.][]{Kormendy95, Magorrian98, Ferrarese00, Tremaine02}. These correlations are thought to be driven by AGN feedback, defined as the impact that growing supermassive black holes have on the host galaxies in which they reside. This impact can happen through radiation, winds, and jets (see \citealt{Harrison24} for a recent review). In cosmological simulations, AGN feedback plays a crucial role in, for example, quenching star formation, which limits the number of massive galaxies \citep{Dubois16}. Furthermore, the energy released by the AGN affects the fueling of the SMBH itself, possibly regulating the AGN duty cycle \citep[][]{Ineson15, Olivares22}. AGN are thought to be a phase of limited duration \citep[$\sim$0.1--100 Myr;][]{Martini04, Novak11, Hickox14} that might occur in every galaxy hosting a SMBH, provided that gas is supplied to the innermost regions of galaxies. This gas can be provided by galaxy mergers, secular processes, and/or direct gas accretion from filaments. Still, regardless of the mechanism, the gas has to lose most of its angular momentum to reach the sphere of influence of the SMBH. This can happen through nuclear bars and mini-spirals, as shown by high angular resolution observations of molecular gas in nearby AGN \citep{Combes03, Garcia-Burillo05, Garcia-Burillo12, Davies09, Hicks09, Audibert19, Kolcu23, Bianchin24}.

At low redshifts, molecular gas is the material from which stars form, constituting a significant fraction of the interstellar medium (ISM). In the local universe, using near-infrared observations, it is possible to observe $\rm{H_2}$ emission lines associated with rovibrational transitions that probe molecular gas at temperatures of $\rm{T>1000\,K}$. Their ratios permit to discriminate between thermal and non-thermal excitation, as well as determining the thermal excitation temperature \citep{Davies04, Davies05, RodriguezArdila05, RamosAlmeida09}. The problem is that the fraction of gas at these high temperatures is typically very low compared with the total molecular gas mass \citep[hot-to-cold gas mass ratios of 10$^{-7}$--10$^{-5}$ for nearby star-forming galaxies;][]{Dale05, Davies06, Pereira-Santaella18}. For this reason, most studies use the transitions of carbon monoxide (CO) and hydroxyl (OH) present in the (sub)-millimeter and far-infrared as indirect tracers of molecular gas. The CO emission can be accessed, for example, with the Atacama Large Millimeter/submillimeter Array (ALMA) and Northern Extended Millimetre Array (NOEMA) observatories and corresponds to gas at temperatures of approximately 5$-$20 K for disks and 50$-$75 K for nuclear regions \citep[see][for reviews]{Cazaux02, Dale02, Habart05}. Meanwhile, observations of the absorption lines of OH at 119 $\mu$m and 163 $\mu$m in nearby galaxies were obtained with the Herschel Space Observatory \citep{Sturm11, Stone16, Runco20}. The measured intensities through these indirect molecular gas tracers can then be converted to H$_2$ column densities using a factor, $\rm{\alpha_{CO}}$ or $X_{OH}$, which are determined under the assumption of identical physical conditions (temperature and density) in all clouds. As a result, they can vary by a factor of $\sim$10 depending on whether the gas is optically thin or exposed to Galactic excitation conditions (see \citealt{Herrera-Camus20} and references therein). With the advent of space telescopes such as the Infrared Space Observatory (ISO) and Spitzer Space Telescope, the pure rotational H$_2$ emission lines became observable in the mid-infrared \citep{Roussel07, Pereira-Santaella13, Pereira-Santaella14}. These lines trace molecular gas at temperatures between $\sim$100 and 1000 K, commonly known as warm molecular gas. By combining the information they provide with CO, OH, and the hot H$_2$ gas, it becomes possible to obtain a complete and comprehensive picture of molecular gas emission in galaxies and to study their physical properties, including temperature, density, column density, etc. \citep[e.g.][]{Davies06}.

The integral field capabilities of the JWST \citep[][]{Gardner23}, combined with ALMA, now offer the opportunity to study the molecular phase of the ISM in galaxies with unprecedented sensitivity, spectral, and spatial resolution, making it possible to characterize the molecular gas kinematics in a spatially-resolved manner. In the case of nearby AGN, this permits us to advance our understanding of the fueling and feedback mechanisms \citep[e.g.][]{Pereira-Santaella22, Garcia-Bernete22, Bianchin24, Davies24, Garcia-Burillo21, Zhang23, Garcia-Burillo24, Hermosa-Munoz24}.

In this work, we target the nearby active galaxy MCG$-$05$-$23$-$16, which is part of the Galactic Activity, Torus, and Outflow Survey (\href{https://gatos.myportfolio.com/}{GATOS}). The aim of this collaboration is to understand the gas flow cycle of local AGN, the properties and evolution of the obscuring material, and the relationship between star formation and nuclear activity in a nearly complete sample of AGN \citep{Garcia-Burillo21, Garcia-Burillo24, Alonso-Herrero21, Garcia-Bernete24a, Garcia-Bernete24b, Zhang24b}.

\begin{figure*}
\centering
\includegraphics[width=1.6\columnwidth, trim={0 0 0 0},clip]{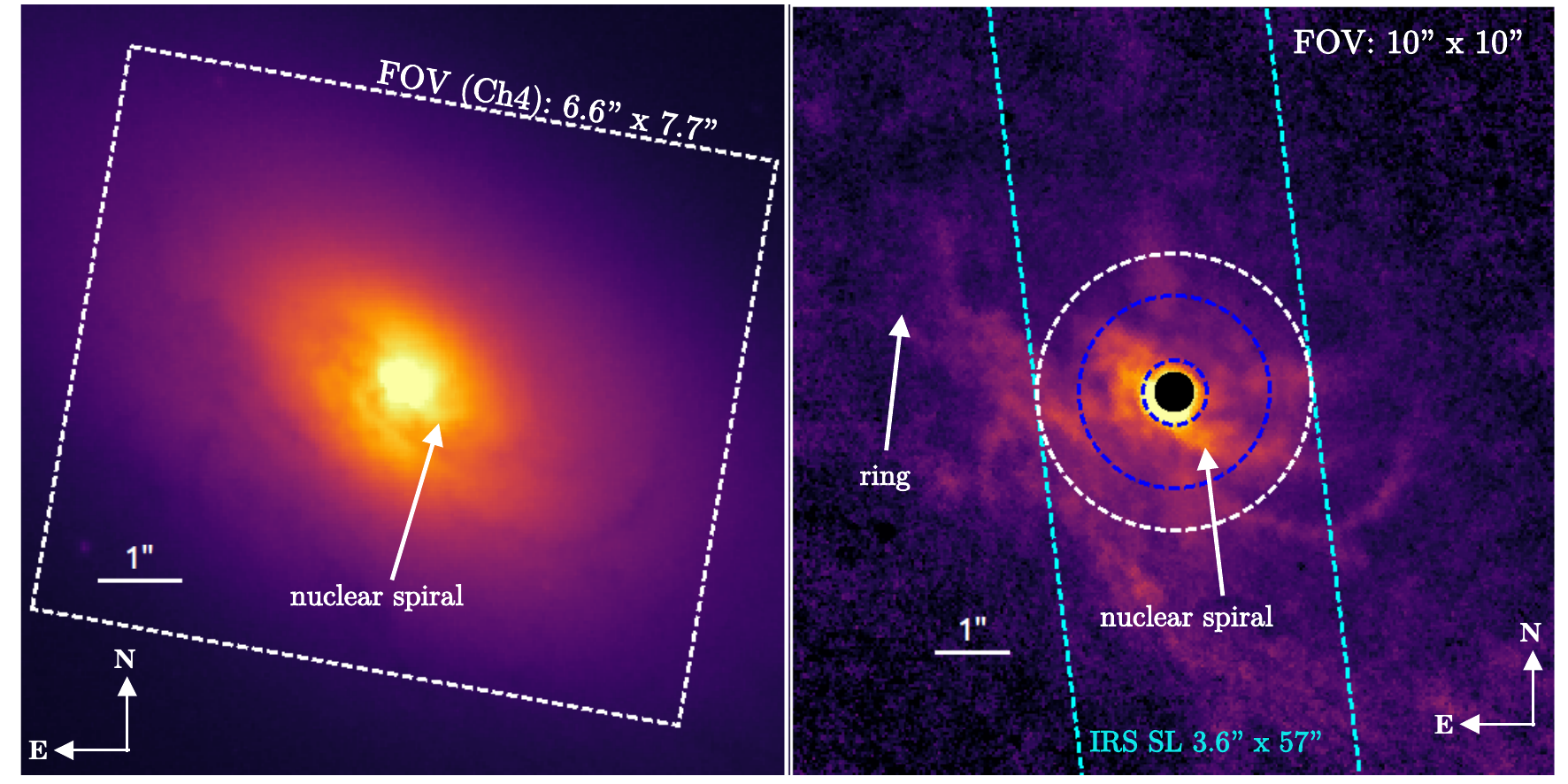}
\label{fig:MCG-05-23-16_HST}
\caption{HST/WFC3 image and color map of MCG$-$05$-$23$-$16. Left: 10\arcsec~$\times$~10\arcsec~image obtained using the filter F606W ($\rm{\lambda_{ref} \sim 5889}$\,\AA). The JWST/MIRI/MRS FOV of Channel 4 (Ch4) is shown with a white dashed rectangle. The horizontal white line indicates the physical size of 1\arcsec. Right: V-H color map using the F606W (left panel) and F160W ($\rm{\lambda_{ref} \sim 15369}$\,\AA) images. The innermost region of the map has been masked using a black solid circle of 1.5 $\times$ FWHM of the F160W filter (1.5 $\times$ 0\farcs18) to avoid artifacts due to the mismatch between the different PSFs of F606W and F160W. The blue dashed circles correspond to the regions used to extract the JWST/MIRI nuclear spectrum considering the correction per aperture (Ch1-s: diameter of 0\farcs8 $\simeq$ 141pc - Ch4-l: 2\farcs5$\simeq$440pc). The white dashed circle corresponds to the region that extracts the extended (diameter of 3\farcs6 $\simeq$ 634 pc) JWST/MIRI spectrum. The nuclear and extended spectra are shown in Fig. \ref{figure:MIRISpectra} with orange and blue solid lines, respectively. The Spitzer/IRS SL (3\farcs6$\times$57\arcsec, PA=187$\degr$) slit is shown with dashed cyan lines. A dust lane resembling a nuclear spiral can be seen southwest and northeast of the nucleus. See also Fig. \ref{Fig:ALMA_vs_HST} in Appendix \ref{app:ALMAvsHST}.}
\end{figure*}
\noindent

MCG$-$05$-$23$-$16 is an spheroidal galaxy (S0) hosting a Seyfert 1.9 nucleus \citep{Veron-Cetty06,Davies20}, with broad recombination lines detected in the near-infrared \citep{Goodrich94} and a bolometric luminosity of $\log (L_{\rm{bol}})\, \rm{[erg\, s^{-1}]}\, \approx 44.3$ \citep{Davies20}. This source has been widely studied in X-rays due to its continuum variability \citep[e.g.][]{Reeves07, Zoghbi14, Marinucci22, Liu24}. The galaxy has also been observed at optical wavelengths with the Hubble Space Telescope (HST; see Fig. \ref{fig:MCG-05-23-16_HST}). Previous studies report that its [O {\small III}]\,5007\AA~and H$_{\alpha}$ emission are elongated on either side of the nucleus with a position angle (PA) of $40^{\circ}$ \citep[e.g.][]{Ferruit00, Prieto14}. \cite{Ruschel-Dutra21} explored the [OIII] and [NII] emission and kinematics using IFU data from the Gemini Multi-Object Spectrographs (GMOS) and reported a kinematic major axis PA=66$^{\circ}$. They did not find clear signatures of ionized gas outflows for this source. More recently, using the same JWST dataset employed here, \cite{Zhang24a} reported tentative evidence of ionized gas outflows, especially along the direction of their AGN ionization cones (PA = 172º) with velocities $> 100$ km/s in this target.

Previous studies using mid-infrared images of the continuum emission from ground-based telescopes classified this source as spatially unresolved \citep[e.g.][]{Asmus15}, and reported good agreement between the ground- and space-based photometric and spectroscopic observations at these wavelengths (see \href{http://dc.zah.uni-heidelberg.de/sasmirala/q/prod/qp/MCG-5-23-16}{Sasmirala} object pages). This indicates that no significant flux or spectral changes happened at these wavelengths over a period of $\sim$25 years, and also that the mid-infrared emission is dominated by the AGN. Studies at radio frequencies found weak diffuse emission at 8.4 GHz with the Very Large Array (VLA), which is marginally resolved with PA $\sim \, 169^{\circ}$ and associated with a compact radio jet \citep{Mundell09, Orienti10}. This paper reports and analyzes the distribution and kinematics of the warm and cold molecular gas in MCG$-$05$-$23$-$16. Using JWST and ALMA together for the first time, we study the H$_2$ pure rotational emission lines and the CO(2$-$1) emission line at the rest frequency of 230.538 GHz. Throughout this work, we assume a cosmology with $\rm{H_0 = 70 \, km \, s^{-1} \, Mpc^{-1}}$, $\Omega_{M} = 0.27$, and $\rm{\Omega_{\Lambda} = 0.73}$. The redshift of the galaxy is z=0.008486, which corresponds to a spatial scale of 176 pc/arcsec and a luminosity distance of 37 Mpc \citep{Kawamuro16}.

\section{Observations and data reduction}
\label{sec:Observations}

\subsection{JWST/MIRI observations}
Observations of MCG$-$05$-$23$-$16 were obtained as part of the Cycle 1 GO program 1670 (PIs: T. Shimizu and R. Davies) using the Medium-Resolution spectrometer (MRS) mode of the Mid-Infrared Instrument \citep[MIRI, ][]{Rieke15, Wright15, Wright23} onboard JWST. MCG$-$05$-$23$-$16 was observed on 3 April 2023, with a single pointing and an on-source exposure time of 1121\,s in each of the four MRS sub-bands using the FASTR1 read-out mode and 4-point dither pattern. Background observations of half the exposure time of the science observations were also obtained using a 2-point dither pattern.

The MIRI/MRS data covers the spectral range 4.9$-$28.1 $\rm{\mu m}$, with a spectral resolution of R $\sim$ 3700$-$1300 \citep{Labiano21,Argyriou23}. The data are split into four channels that cover the following ranges: 4.9$-$7.65 $\mu$m (Channel 1; hereafter Ch1), 7.51$-$11.71 $\mu$m (Channel 2; Ch2), 11.55$-$18.02 $\mu$m (Channel 3; Ch3), and 17.71$-$28.1 $\mu$m (Channel 4; Ch4). Combining these four channels and the three grating settings available (short, medium, and long), there are 12 different wavelength bands. We measured the angular resolutions for the 12 bands from the full-width at half maximum (FWHM) of the star HD 163466 (program ID: 01050; PI: B. Vandenbussche), observed on 9 June 2022. These are reported in Table \ref{tab:fwhm}, together with the field-of-view (FOV) of each channel.

The data were reduced using the JWST Science Calibration Pipeline (version 1.11.4), with the context 1130 for the calibration References Data System \citep{Labiano16, Alvarez-Marquez23, Bushouse24}. Some hot and cold pixels are not identified by the current pipeline version, so we added an extra step before creating the data cubes to mask them. The data reduction is described in detail in \cite{Garcia-Bernete22, Garcia-Bernete24a} and \cite{Pereira-Santaella22}. The background subtraction was done spectrum by spectrum for each science cube using background average spectra. These average spectra were generated through cubes created with the dedicated background-only observations of each sub-band for each channel. We also applied a residual fringing correction to each spectrum of the science cubes using the script {\sc rfc1d-utils} provided by the JWST help desk \citep{Argyriou23}. This \href{https://jwst-pipeline.readthedocs.io/en/latest/jwst/extract_1d/description.html}{correction} is now available in the latest pipeline version.

\begin{table}
\caption{FWHM and FOV of MIRI/MRS channels and bands.}
\label{tab:fwhm}
\centering
\begin{footnotesize}
\begin{tabular}{lccc} 
 \hline
 MRS channel & PSF FWHM & FOV & H$_2$ lines \\
 \& band     & (arcsec) & (arcsec) & detected \\
 \hline\hline
Ch1-short (Ch1-s) & 0.278 & 3.2$\times$3.7 & \dots\\
Ch1-medium (Ch1-m) & 0.290 & 3.2$\times$3.7 & \dots\\
Ch1-long (Ch1-l) & 0.294  & 3.2$\times$3.7 & 0$-$0S(5)\\
\hline
Ch2-short (Ch2-s) & 0.359 & 4.0$\times$4.8 & 0$-$0S(4)\\
Ch2-medium (Ch2-m) & 0.379 & 4.0$\times$4.8 & 0$-$0S(3)\\
Ch2-long (Ch2-l) & 0.423  &  4.0$\times$4.8 & \dots\\
\hline
Ch3-short (Ch3-s) & 0.503 &  5.2$\times$6.2 & 0$-$0S(2) \\ 
Ch3-medium (Ch3-m) & 0.541 & 5.2$\times$6.2 & \dots\\
Ch3-long (Ch3-l) & 0.603 &   5.2$\times$6.2 & 0$-$0S(1) \\
\hline
Ch4-short (Ch4-s) & 0.760 & 6.6$\times$7.7  & \dots\\
Ch4-medium (Ch4-m) & 0.823 & 6.6$\times$7.7  &\dots \\
Ch4-long (Ch4-l) & 0.835   &6.6$\times$7.7  &\dots \\
\hline
\end{tabular}
\end{footnotesize}
\tablefoot{FWHM measured from the collapsed spectrum of the star HD 163466 in each of the 12 sub-channels of MIRI/MRS, and FOV of each channel. The pixel sizes employed to convert from pixels to arcsec are 0\farcs13~in Ch1, 0\farcs17~in Ch2, 0\farcs20~in Ch3, and 0\farcs20~in Ch3, 0\farcs35~in Ch4 (see \href{https://jwst-docs.stsci.edu/jwst-mid-infrared-instrument/miri-observing-modes/miri-medium-resolution-spectroscopy\#gsc.tab=0}{MIRI Observations Modes}). The last column indicates the channels where we detect the H$_2$ lines from S(5) to S(1), which have rest-frame wavelengths of 6.90952, 8.02505, 9.66492, 12.27861, and 17.03484 $\mu$m, respectively.}
\end{table}

\subsection{ALMA and VLA data}
\label{sec:alma}

ALMA observations of MCG$-$05$-$23$-$16 in band 6 were performed in two execution blocks (project ID 2019.1.01742.S, PI: D. Rosario) for a total on-source integration time of 15.6 minutes in two configurations: compact (TC) and extended (TE). The TE configuration (baselines from 15 to 2516\,m) was observed with 50 antennas in October 2019 and an on-source integration time of 11.6 minutes, resulting in a synthesized beam of 0\farcs39~($\sim$70 pc). The TC configuration (baselines from 15 to 314\,m) was observed in November 2019 with 47 antennas and an on-source integration time of 4 minutes, providing a synthesized beam of 1\farcs44~($\sim$250 pc). The observations were centered on the nucleus, with a single pointing covering a field of view of 26\arcsec. The correlator setup was centered at the observed frequency of the CO(2$-$1) line: 228.597\,GHz, with a bandwidth of 1.875\,GHz and channel spacing of 7.8\,MHz ($\sim$10.2\,km~s$^{-1}$). The phase calibrator J1007$-$3333 was used both in TE and TC, and J1037$-$2934 was used as flux calibrator in TE configuration and J0725$-$0054 in TC, from which we measured the typical 10\% flux accuracy.

The data were calibrated, concatenated, and cleaned with the \textsc{casa} software (version 5.6.1-8; \citealt{McMullin07}). Once each configuration was calibrated, we concatenated the two of them using the \textsc{concat} task. The imaging and cleaning were performed with the
task \textsc{tclean}. The spectral line map was obtained after subtracting the continuum in the uv-plane using the task \textsc{uvcontsub} specifying a zeroth-order polynomial in the channels free from emission lines. The CO(2$-$1) data cube was produced with a spectral resolution of 15\,km~s$^{-1}$ and using Briggs weighting mode \citep{Briggs95} and a robust parameter set to 2 to achieve the best sensitivity, equivalent to a natural weighting. The resulting synthesized beam is 0\farcs65$\times$0\farcs53 at PA of 87$^\circ$ and a rms noise of 0.51\,mJy beam$^{-1}$ per 15\,km~s$^{-1}$ channel. Finally, the datacube was corrected for primary beam attenuation.

We also retrieved the X-band VLA radio image (8.4\,GHz) of our target from the \href{https://www.vla.nrao.edu/astro/nvas/}{NRAO} VLA Archive Survey Images Page, observed on 21 August 1999 in A configuration with a beam size of 0\farcs33~($\sim$60 pc; synthesized beam of 0\farcs43$\times$0\farcs23) and a rms of 0.1 mJy. The data was first presented in \citet{Orienti10}, where the authors reported that the total radio spectrum slightly steepens at 8.4 GHz with a spectral index of 0.8$\pm$0.1.


\section{Analysis and results}
\label{sec:Results}

\subsection{Nuclear and extended mid-infrared spectra}

\begin{figure}
\centering
\includegraphics[width=1.0\columnwidth, trim={5 0 5 0},clip]{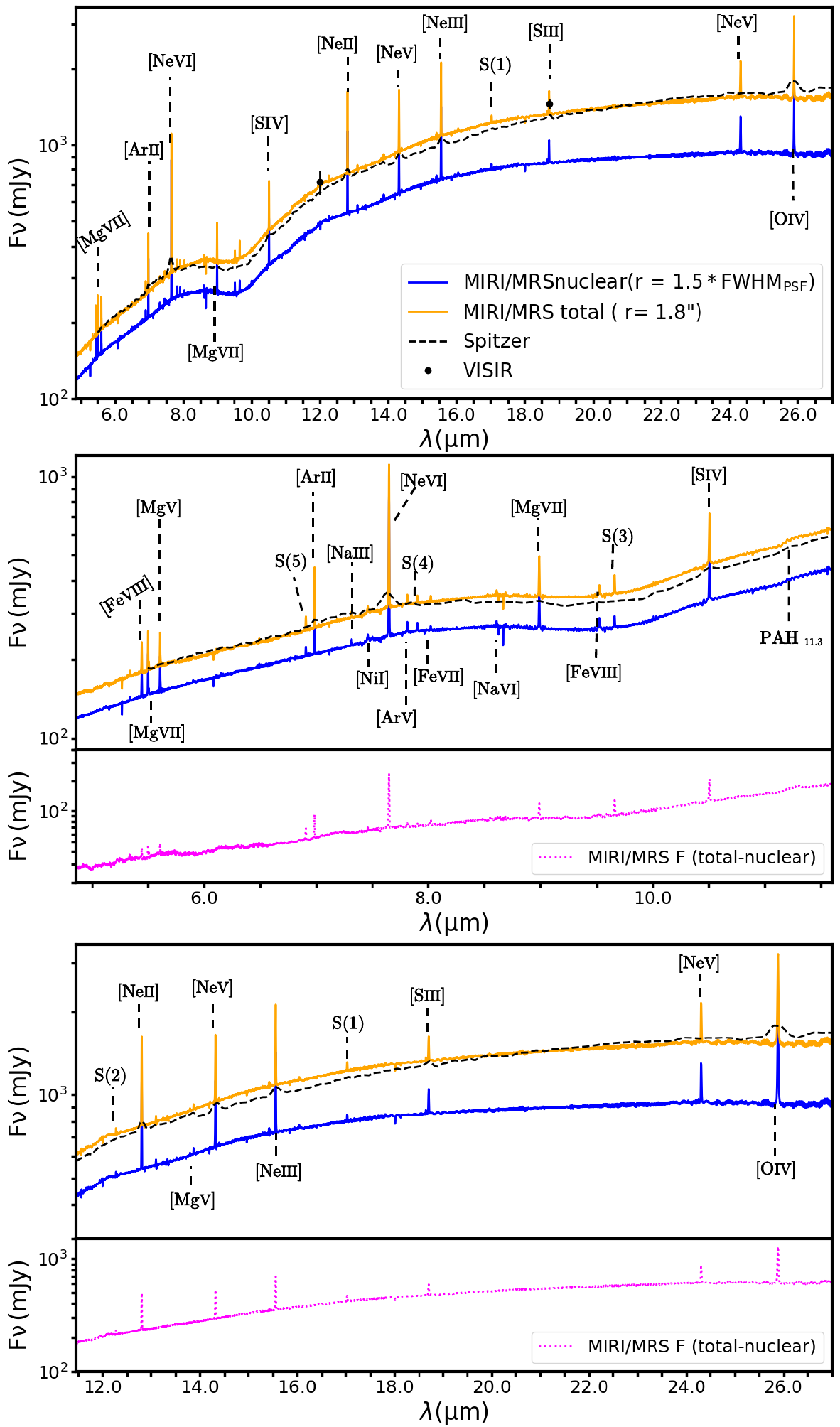}
 \caption{Mid-infrared integrated spectra of MCG$-$05$-$23$-$16. The blue and orange solid lines correspond to the nuclear and extended JWST/MIRI spectra, extracted in circular apertures of $\sim$ 0\farcs8$-$2\farcs5~(146-441 pc) and 3\farcs6~(634 pc) diameter, respectively (black and white circles in Fig. \ref{fig:MCG-05-23-16_HST}). The magenta dotted line represents the spectrum obtained by subtracting the nuclear from the extended spectrum. The dashed black line corresponds to the Spitzer/IRS spectrum, extracted in the same aperture as the extended JWST/MIRI spectrum. The middle and bottom panels show zooms of two spectral regions. The black filled circles mark the nuclear 12$\mu m$ and 18$\mu m$ continuum emission reported by \cite{Asmus15}, using data from VISIR/VLT.}
\label{figure:MIRISpectra}
\end{figure}

The left panel of Fig. \ref{fig:MCG-05-23-16_HST} shows an HST/WFC3 optical image of the central 10\arcsec$\times$10\arcsec~($\sim$1.7$\times$1.7 kpc$^2$) of MCG$-$05$-$23$-$16 (proposal ID: 15181; PI: D. Rosario). The V-H color map using the F606W and F160W images is shown in the right panel.  The blue and white circles show the regions used to extract the nuclear and extended spectra, shown as blue and orange lines in Fig. \ref{figure:MIRISpectra}. The nuclear spectrum was extracted as a point source, with increasing form 1.5 $\times$ PSF FWHM in each channel (see Table \ref{tab:fwhm}.). These radii range from 0\farcs42 on Ch1-s to 1.252 on Ch4-l. The extended spectrum was extracted in an aperture of 1\farcs8~radius centered in the nucleus (diameter of 3\farcs6 $\simeq$ 634 pc) and including the nuclear aperture. The spectra from the four channels were combined using different scaling factors to align them in overlapping areas. These scaling factors were determined by conducting linear fits, and they represent less than 1.2\% of the flux measured at any given wavelength in the final spectra shown in Fig. \ref{figure:MIRISpectra}.

In Fig. \ref{figure:MIRISpectra} we also show the Spitzer/IRS spectra downloaded from CASSIS (Combined Atlas of Sources with Spitzer IRS Spectra; \citealt{Lebouteiller11}). This spectrum was obtained by putting together the spectra from the IRS Short-Low (SL) and Long-Low (LL) slits, which have resolutions of R$\sim$60$-$127 and 57$-$126, respectively. The wavelength range between 5.3 and 14 $\mu$m is obtained through the SL slits, which have a width of $\sim$3\farcs6~(634\,pc), and the 14$-$38 $\mu$m range through the LL slits, of $\sim$10\farcs5~width (1.85 kpc). The size and PA of the SL slit are shown in the right panel of Fig. \ref{fig:MCG-05-23-16_HST}. Furthermore, we include the nuclear $12\mu$m and $18\mu$m continuum emission obtained through VISIR/VLT data and reported by \cite{Asmus15} in \href{http://dc.zah.uni-heidelberg.de/sasmirala/q/prod/qp/MCG-5-23-16}{Sasmirala} catalog.

The continuum shape of the IRS and MIRI extended spectra are almost identical, as expected, since they were extracted in the same aperture. Thanks to the superior spectral resolution and sensitivity of JWST/MIRI, a large number of previously undetected emission lines are observed in both the nuclear and extended MIRI spectra shown in Fig. \ref{figure:MIRISpectra}. In this work, we are particularly interested in the H$_2$ rotational transitions that trace warm molecular gas at hundreds of Kelvin. In the MIRI spectra shown in Fig. \ref{figure:MIRISpectra} we detect H$_2$0$-$0S(1)\,$17.03\,\mu$m, S(2)\,$12.28\,\mu$m, S(3)\,$9.66\,\mu$m, S(4)\,$8.02\,\mu$m, and S(5)\,$6.90\,\mu$m. The more precise rest-frame wavelength of these emission lines and the MRS channel in which they are detected are indicated in Table \ref{tab:fwhm}.

The nuclear spectrum, extracted in a circular aperture of 146-634 pc in diameter, shows the same spectral shape as the extended spectrum, consistent with an AGN-dominated mid-infrared emission. The 10 $\mu$m silicate feature appears in shallow absorption in both the nuclear and extended-spectrum, indicating that the obscuring material producing it comes from the nuclear region. Considering that the silicate feature lies between Ch2-m and Ch2-l, this corresponds to a diameter of 1.33-1\farcs27 (200-223\,pc). The 18 $\mu$m silicate feature is practically absent \citep{Garcia-Bernete24a}.

\subsection{Morphology and kinematics of the molecular gas}

\subsubsection{Warm versus cold molecular gas from pc to kpc scales.}
\label{morphology}

\begin{figure*}
\centering
\includegraphics[width=1.43\columnwidth, trim={0 20 5 0},clip]{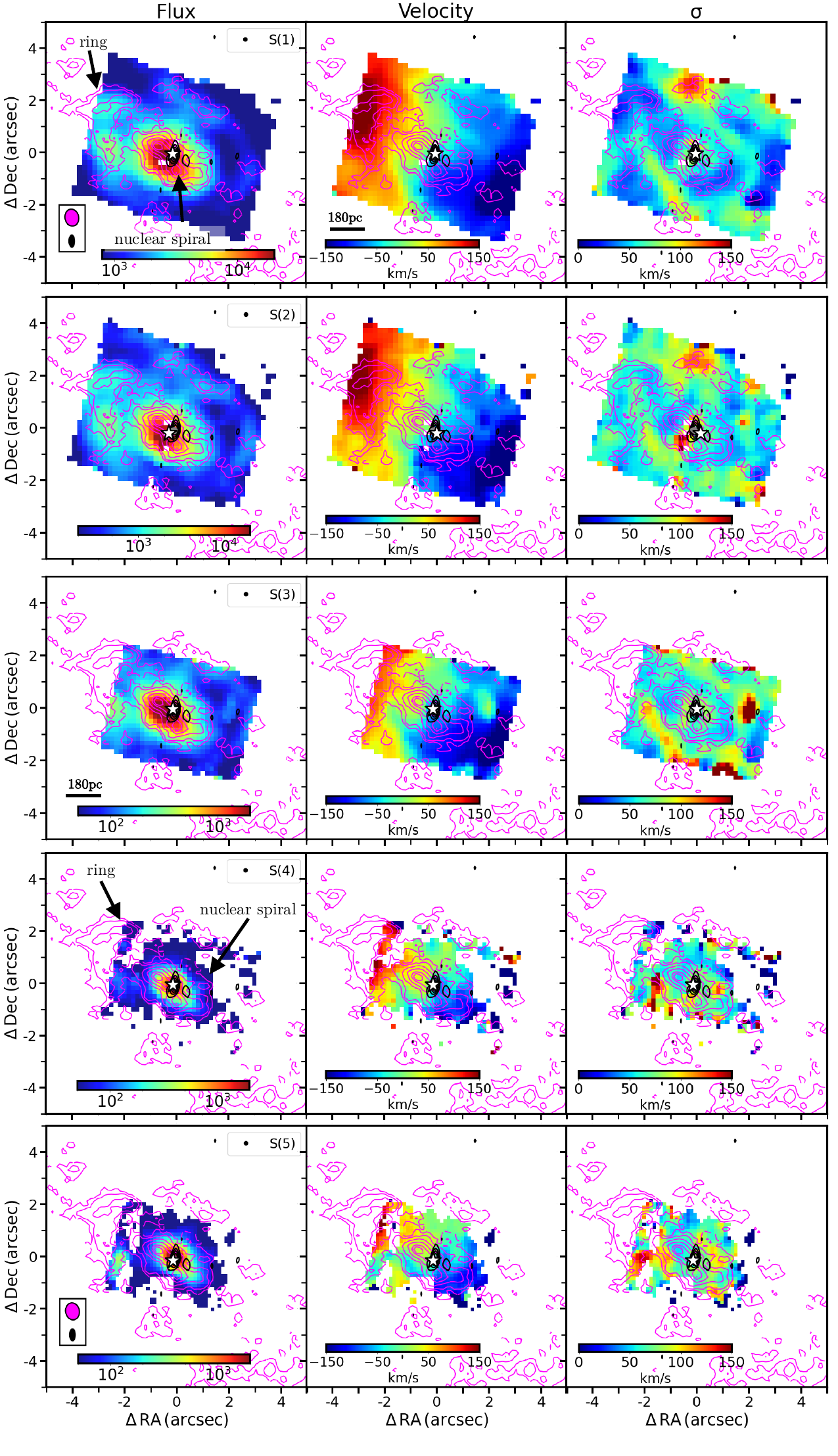}
 \caption{Flux $\rm{[\times 10^{-20}\,erg\,s^{-1}\,cm^{-2}]}$ (left), velocity [$\rm{km/s}$] (center), and velocity dispersion [$\rm{km/s}$] (right) maps of the different $\rm{H_2}$ rotational transitions detected in the MIRI/MRS data. The magenta contours represent 10\%, 20\%, 30\%, 40\%, 50\%, 60\%, 70\%, 80\%, and 90\% of the maximum ALMA CO(2$-$1) flux. For the S(1), S(2), and S(3) maps we also included the ALMA contours corresponding to 7\% and 8\% of the maximum flux. The black contours correspond to the 5\%, 10\%, 15\%, and 20\% of the average $\sigma$ ($\sim 2.85 \times 10^{-5}$) in the VLA 8.4 GHz image. The white star shows the AGN location according to the ALMA continuum. The VLA and ALMA beams, corresponding to 0\farcs14 $\times$ 0\farcs30 and 0\farcs68 $\times$ 0\farcs83, are shown in the S(1) and S(5) flux maps.}
\label{fig:Maps_H2}
\end{figure*}

To understand the behavior of the molecular gas, we produced kinematic maps of the different $\rm{H_2}$ lines using the cubes corresponding to Ch3-l, Ch3-s, Ch2-m, Ch2-s, and Ch1-l (see Table \ref{tab:fwhm}) and the \textsc{alucine} tool \citep{PeraltadeArriba23}. This tool creates flux, velocity, and velocity dispersion ($\sigma$) maps by fitting the emission lines using single or multiple Gaussian components and a local continuum on a spaxel-by-spaxel basis. The flux, velocity, and velocity dispersion maps of the five rotational transitions of the $\rm{H_2}$ molecule considered here are shown in Fig. \ref{fig:Maps_H2}. We used a single Gaussian component to fit the emission line profiles. Since MCG$-$05$-$23$-$16 is a Seyfert 1.9 galaxy, the mid-infrared continuum dominates the central region of the cubes, making the equivalent width (EW) of the emission lines very low. This implies that some of the central spaxels show negative values when we subtract the continuum. To minimize this effect, we applied spatial smoothing to the cubes before running \textsc{alucine} using the \emph{smooth} tool of \textsc{qfitsview}. We selected a boxcar average of one pixel along both the X and Y axes. After applying this smoothing, only a few spaxels with negative values remained in the $\rm{H_2}$0$-$0S(1) and S(2) maps, that is, those corresponding to Ch3 (see the first and second rows of Fig. \ref{fig:Maps_H2}).

Unlike the ionized gas traced by the various emission lines detected in the MIRI/MRS spectrum (see Fig. \ref{figure:MIRISpectra}), which is dominated by the nuclear emission (see \citealt{Zhang24a} for more details on the ionized gas kinematics), the warm molecular gas is extended over the entire FOV of the different channels. For comparison, in Fig. \ref{fig:Maps_H2}, we plot the contours of the CO(2$-$1) emission obtained from the ALMA data (see moment maps in Fig. \ref{Fig:MomentMapCO} of Appendix \ref{app:moments}) in magenta solid line and the radio continuum emission at 8.4 GHz from VLA in black solid line (see Sect. \ref{sec:alma} for a description of the ALMA and VLA observations). We find that the morphologies of the CO(2$-$1) and H$_2$ emission are similar, centrally peaked, and resembling a mini-spiral in the central $\sim$2\arcsec~($\sim$350 pc) that coincides with the dust lane seen in the HST optical image of the galaxy (see Fig. \ref{fig:MCG-05-23-16_HST}).
This mini-spiral is connected with a ring-like feature of almost the same size as the FOV of Ch3 ($\sim$8\arcsec$\simeq$1.4 kpc). The northeast side of the ring is the most conspicuous in H$_2$, as it can be seen from the S(1) and S(2) flux maps shown in Fig. \ref{fig:Maps_H2}.

Although Ch1, Ch2, and Ch3 cover different FOVs (see Table \ref{tab:fwhm}), the colder H$_2$ gas traced by the S(1) and S(2) lines (two top rows of Fig. \ref{fig:Maps_H2}) seems to exhibit a more regular rotation pattern than the warmer H$_2$ gas traced by S(3), S(4), and S(5). If we subtract the S(5) from the S(1) velocity map, we find residual velocities of up to $\sim$100 km/s at certain locations, indicating important differences between the kinematics traced by the two lines. The rotation velocities reach $\pm$250 km~s$^{-1}$, redshifted to the northeast and blueshifted to the southwest (see the middle column of Fig. \ref{fig:Maps_H2}). These velocities are similar to the low ionization lines (e.g. [Ne{\small II}]) as shown \cite{Zhang24a} for this source, which is consistent with the kinematics of the warm ionized gas traced by the optical [O {\small III}] and [N {\small II}] emission lines \citep{Ruschel-Dutra21}. The dispersion maps shown in the last column of the same figure show interesting structures, including knots of high-velocity dispersion that reach values of up to 160-170 km~s$^{-1}$. These knots are mainly located in the regions between the ring and the nuclear spiral. An example is the S(3) knot at $\sim$2\arcsec~$\sim$350 pc westward of the nucleus. These areas of high-velocity dispersion coincide with regions that appear depleted of CO(2$-$1). The VLA 8.4 GHz contours show extended emission to the north, perpendicular to the mini-spiral (PA$\sim 170^{\circ}$ and size $\rm{\sim 200\, pc}$). This slightly extended emission was interpreted as a compact radio jet by \cite{Orienti10}, but it does not seem connected with any of the kinematic features detected in the H$_2$ gas. In Sect. \ref{sec:barolo}, we compare the kinematics of the warm and cold molecular gas in more detail. 

\subsubsection{$\rm{H_2}$ molecular gas in specific regions}
\label{subsec:CommonRegions}

\begin{figure*}
\centering
\includegraphics[width=1.5\columnwidth, trim={0 0 0 0},clip]{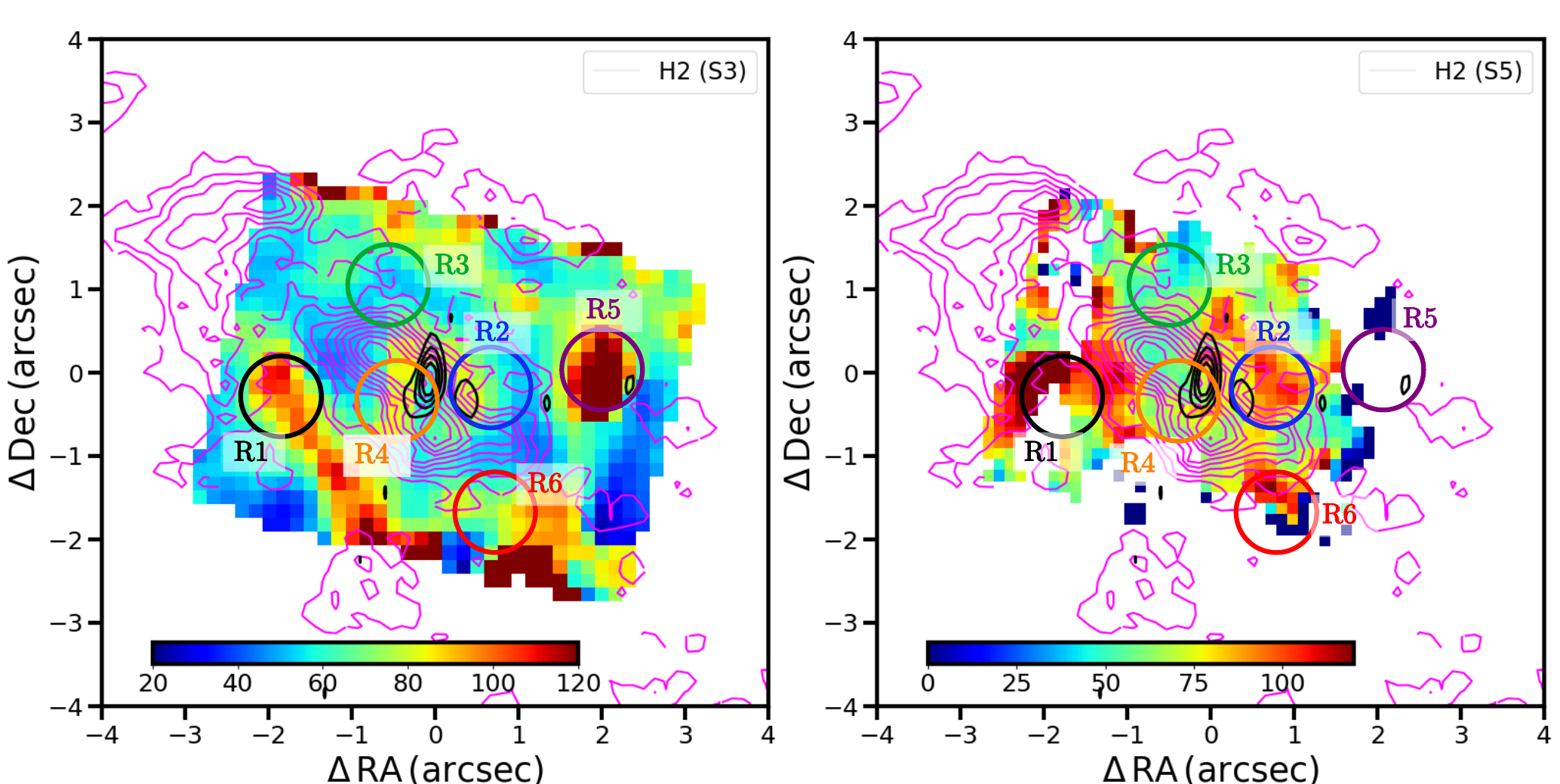}
 \caption{Velocity dispersion maps of the $\rm{H_2}$0$-$0S(3) and S(5) emission lines with the CO(2$-$1) moment 0 contours superimposed. The units of the color bars are km s$^{-1}$. The circles of different colors indicate the regions from where we extracted the spectra shown in Fig. \ref{fig:Profiles_similarRegions_norm}. These regions have radii of 0\farcs5~(88 pc). In Sect. \ref{subsec:CommonRegions}, we describe how the regions were selected.}
\label{fig:Maps_with_regions}
\end{figure*}

This section reports the results from comparing the $\rm{H_2}$ emission line profiles extracted in six regions of 0\farcs5~radius ($\sim$90 pc) corresponding to areas where high-velocity dispersion values are visually identified for one or several $\rm{H_2}$ lines. In Fig. \ref{fig:Maps_with_regions}, we identify these regions with circles of different colors in the S(3) and S(5) velocity dispersion maps. Regions R2 and R4 are the most nuclear ones, with R4 encompassing the nucleus and R2 located westward from it. R3 and R6 are situated northeast and southwest of the nucleus, respectively, and regions R1 and R5 to the east and west of the nucleus, with R5 exhibiting high dispersion in S(3), but the S(5) emission falls outside of the FOV in Ch1-l.

\begin{figure*}
\centering
\includegraphics[width=1.5\columnwidth, trim={5 30 25 10},clip]{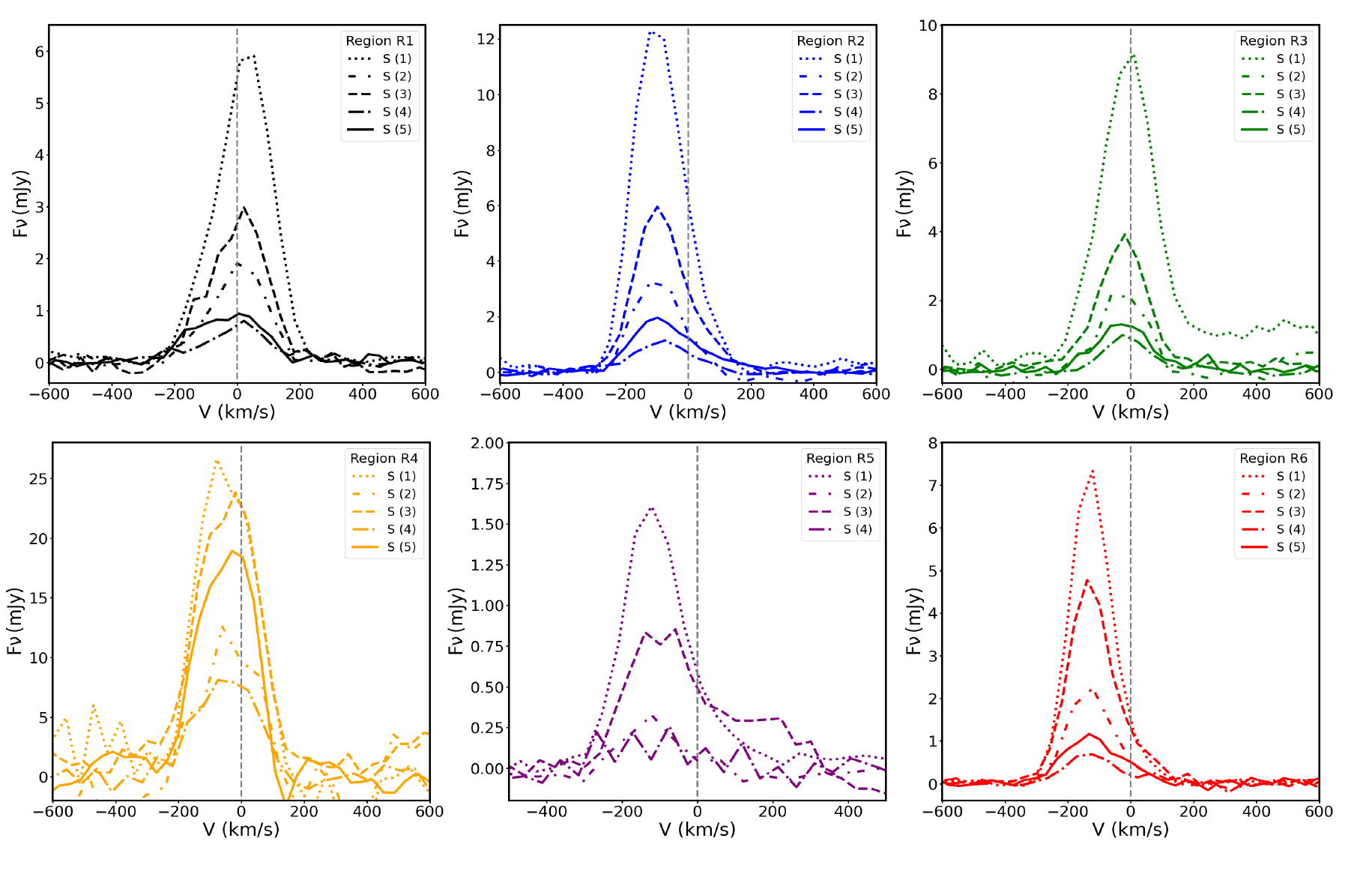}
 \caption{Emission line profiles of the five $\rm{H_2}$ emission lines extracted from the regions with matching colors indicated in Fig. \ref{fig:Maps_with_regions}. All the velocities refer to the systemic value computed with z = 0.008486, indicated by the dashed vertical black lines.}
\label{fig:Profiles_similarRegions_norm}
\end{figure*}

In Figure \ref{fig:Profiles_similarRegions_norm} we show the corresponding emission line profiles extracted from each region. We modeled each line profile with a single Gaussian and subtracted a local continuum, defined by interpolating between two regions adjacent to the emission line. In Table \ref{Tab:H2_emission_SameR}, we report the flux, velocity shift ($\Delta V$), and intrinsic velocity dispersion ($\sigma$) values measured for each line. The integrated flux values of the $\rm{H_2}$ lines in different regions range [0.36:62] $\rm{\times 10^{-16} \, erg \, s^{-1} \, cm^{-2}}$. The highest flux values are found in the region R4, which includes the nucleus. Notably, in this region, the S(3) and S(5) lines show the highest fluxes, while the lowest are measured for S(2) and S(4). The region R5, centered at the position of the S(3) high-velocity dispersion knot at $\sim$2.5\arcsec~westwards of the nucleus, shows the lowest flux values ($\leq$2.5 $\times \rm{10^{-16} \, erg \, s^{-1} \, cm^{-2}}$). When comparing the flux values of the five $\rm{H_2}$ lines in each region, the S(1) emission line is the most luminous in regions R1, R2, and R3, and in R5 and R6, the S(3) flux values are maximum.

We also found that the emission line profiles of different $\rm{H_2}$ emission lines show different shapes (see Fig. \ref{fig:Profiles_similarRegions_norm}). The lines are blueshifted relative to systemic in regions R2, R4, R5, and R6. The region R6, located southwest of the nucleus, shows the highest velocities, |$\rm{\Delta V| > 120 \, \rm{km~s^{-1}}}$. The maximum blueshift is observed for the S(3) emission in this region ($-$139 $\rm{km~s^{-1}}$). In several radial profiles of these lines, we also observe the presence of wings. The S(1) and S(3) lines in region R5 show red wings (see also S(1) in region R3). We observe slightly blue-shifted wings in the region R1 for all transitions, except by the radial profiles of the S(1). In regions R1 and R3, located to the east and northeast of the nucleus, we measure velocities closer to zero, some blueshifted and others redshifted, for all transitions (see Table \ref{Tab:H2_emission_SameR}). Finally, the velocity dispersion values, corrected for the instrument broadening, range between 40 and 170 $\rm{km~s^{-1}}$, increasing from S(1) to S(5) in all regions. The $\rm{\sigma}$ values measured for S(1) are $\leq$64 $\rm{km~s^{-1}}$, while for S(5) are $\geq$140 $\rm{km~s^{-1}}$ in all regions (see Table \ref{Tab:H2_emission_SameR}). This suggests that the S(1) and S(2) lines, which probe colder molecular gas, may trace gas that is more rotation-dominated, while the S(3) and S(5) lines may be associated with more dispersion-dominated gas.

\begin{table*}
    \caption{Flux, velocity, and velocity dispersion values of the $\rm{H_2}$ lines in the integrated regions ($\rm{r = }$0\farcs5) indicated in Fig. \ref{fig:Maps_with_regions}.}
        \label{Tab:H2_emission_SameR}
        \centering
        \begin{scriptsize}
        \setlength{\tabcolsep}{2pt}
        \renewcommand{\arraystretch}{1}
           \begin{tabular}{c|ccc|ccc|ccc|ccc|ccc|ccc}
             \hline \hline
            $\rm{H_2}$ & \multicolumn{3}{|c|}{Region R1} & \multicolumn{3}{|c|}{Region R2} & \multicolumn{3}{|c|}{Region R3} & \multicolumn{3}{|c|}{Region R4} & \multicolumn{3}{|c|}{Region R5} & \multicolumn{3}{|c}{Region R6} \\            
             & Flux & $\rm{\Delta V}$ & $\sigma$ & Flux & $\rm{\Delta V}$ & $\sigma$ & Flux & $\rm{\Delta V}$ & $\sigma$ & Flux & $\rm{\Delta V}$ & $\sigma$ & Flux & $\rm{\Delta V}$ & $\sigma$ & Flux & $\rm{\Delta V}$ & $\sigma$ \\
            \hline
            0$-$0S(1)	& $7.67 \pm 0.78$ & 10 & $62 \pm 10$ & $14.94 \pm 1.50$ & $-$78 & $56 \pm 10$ & $11.99 \pm 1.21$ & 10 & $64 \pm 10$ & $36.31 \pm 3.64$ & $-$34 & $64 \pm 10$ & $1.91 \pm 0.20$ & $-$121 & $57 \pm 31$ & $7.12 \pm 0.72$ & $-$121 & $41 \pm 19$ \\ 
            0$-$0S(2)	& $3.32 \pm 0.35$ & 0 & $87 \pm 6$ & $5.31 \pm 0.54$ & $-$121 & $76 \pm 6$ & $3.21 \pm 0.33$ & 0 & $66 \pm 6$ & $20.82 \pm 2.09$ & $-$60  & $ <  41$ & $0.36 \pm 0.12$ & $-$121  & $ <  84$ & $3.27 \pm 0.34$ & $-$121 & $71 \pm 6$ \\ 
            0$-$0S(3)	& $6.64 \pm 0.68$ & 21 & $120 \pm 4$ & $12.43 \pm 1.26$ & $-$99 & $87 \pm 25$ & $7.54 \pm 0.77$ & $-$19 & $100 \pm 4$ & $61.55 \pm 6.17$ & $-$59 & $130 \pm 4$ & $2.46 \pm 0.28$ & $-$59 & $163 \pm 15$ & $8.76 \pm 0.89$ & $-$139 & $95 \pm 4$  \\ 
            0$-$0S(4)	& $2.45 \pm 0.28$ & 22 & $140 \pm 37$ & $3.15 \pm 0.34$ & $-$75 & $112 \pm 29$ & $1.98 \pm 0.21$ & $-$27  & $ <  29$ & $25.18 \pm 2.54$ & $-$27 & $140 \pm 16$ & $0.59 \pm 0.08$ & $-$123 & $<131$ & $1.74 \pm 0.20$ & $-$123 & $94 \pm 27$ \\
            0$-$0S(5) &	$3.91 \pm 0.41$ & $-$29 & $148 \pm 37$ & $6.73 \pm 0.69$ & $-$98 & $139 \pm 30$ & $3.91 \pm 0.41$ & $-$29 & $147 \pm 3$ & $61.98 \pm 6.22$ & $-$64 & $157 \pm 3$ & \dots &\dots&\dots & $3.63 \pm 0.38$ & $-$133 & $170 \pm 3$ \\
            \hline 
            \end{tabular}
            \end{scriptsize}
    \tablefoot{The flux values are in units of $10^{-16}$ $\rm{erg \, s^{-1} \, cm^{-2}}$, and the velocity and velocity dispersion values in $\rm{km~s^{-1}}$. The fluxes have been corrected from extinction by considering the value of A$_V$=1 mag from \citet{Prieto14} and the \citet{Cardelli89} reddening law, and the values of $\sigma$ have been corrected from instrumental width using the values reported in \citet{Argyriou23}.}
\end{table*}

\subsubsection{Modeling the molecular gas kinematics}
\label{sec:barolo}

\begin{figure*}
\centering
\includegraphics[width=1.6\columnwidth, trim={5 5 0 0},clip]{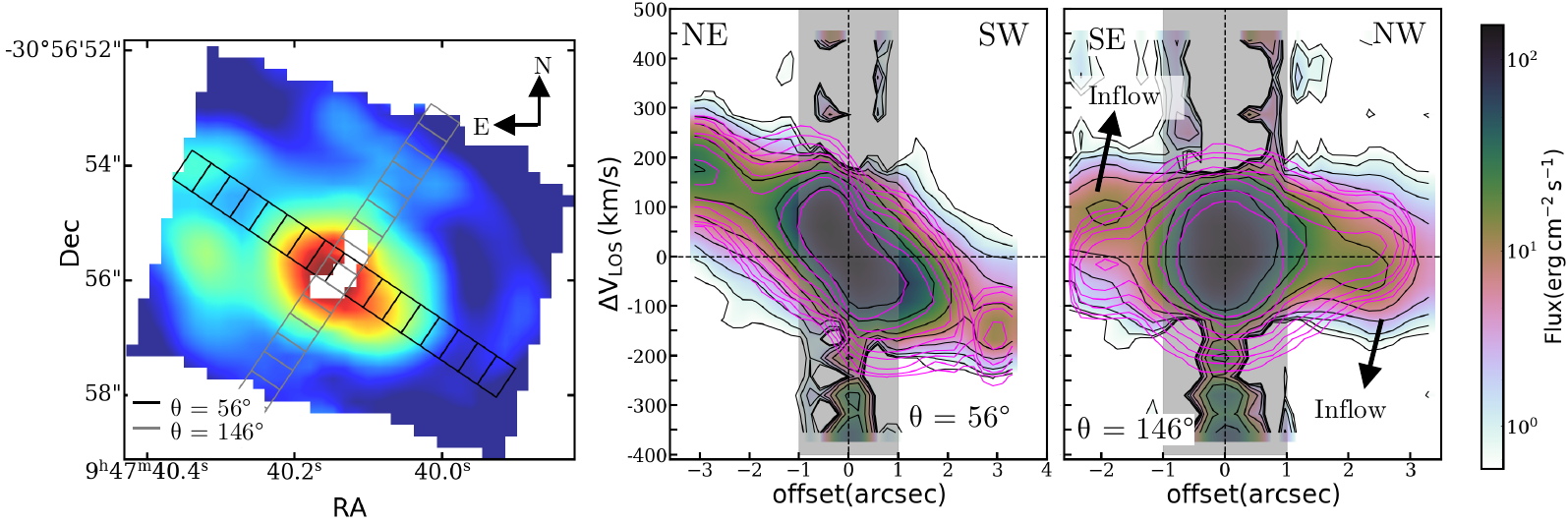}
\includegraphics[width=1.6\columnwidth, trim={5 0 0 0},clip]{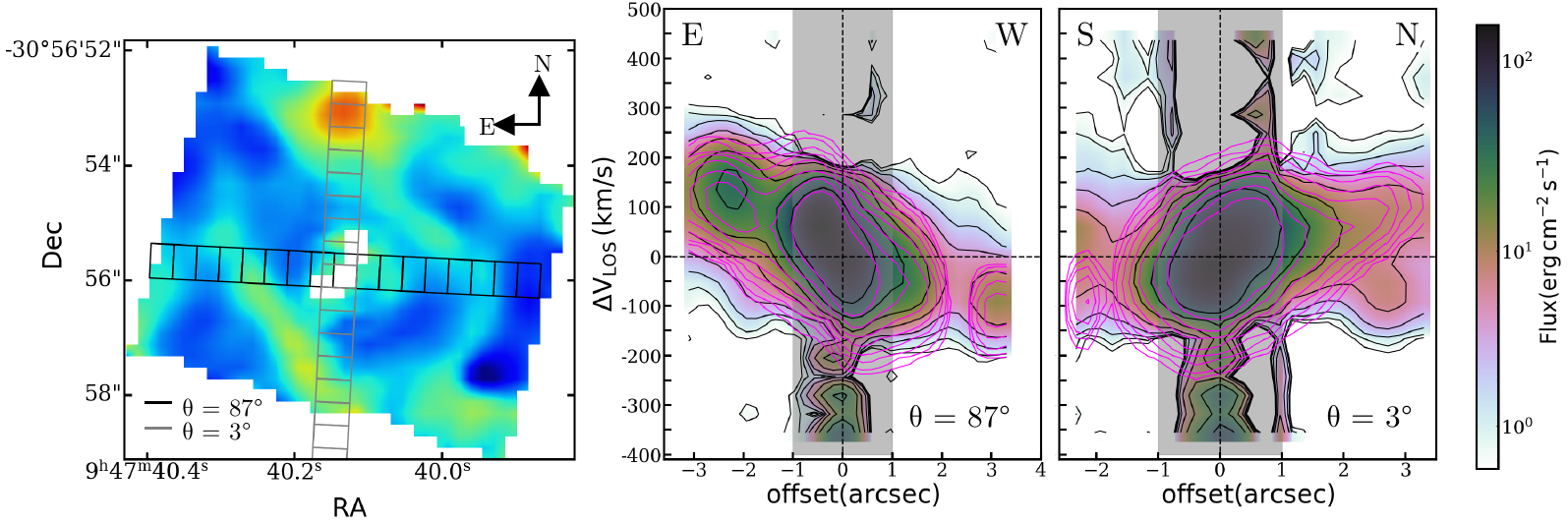}
 \caption{PVDs of the S(1) line. Left panels: Flux and velocity dispersion maps obtained using bilinear interpolation, with the regions used to extract the PVDs superimposed, which have a size of 0\farcs6~$\times$ 7\arcsec. Middle and right panels: PVDs extracted along the kinematic major and minor axes determined by \barolo~(top) and along PA=87$^{\circ}$ and 3$^{\circ}$ (bottom). The black and magenta contours correspond to the data and rotating disk model above 2$\sigma$ (see Sect. \ref{sec:barolo}). The high-velocity emissions observed between -1 and 1 arcsec correspond to artifacts in the nuclear region because of the strong AGN continuum and derived low EW of the molecular emission lines there.}
\label{fig:plots_pvs1}
\end{figure*}

\begin{figure*}
\centering
\includegraphics[width=1.6\columnwidth, trim={5 5 0 0},clip]{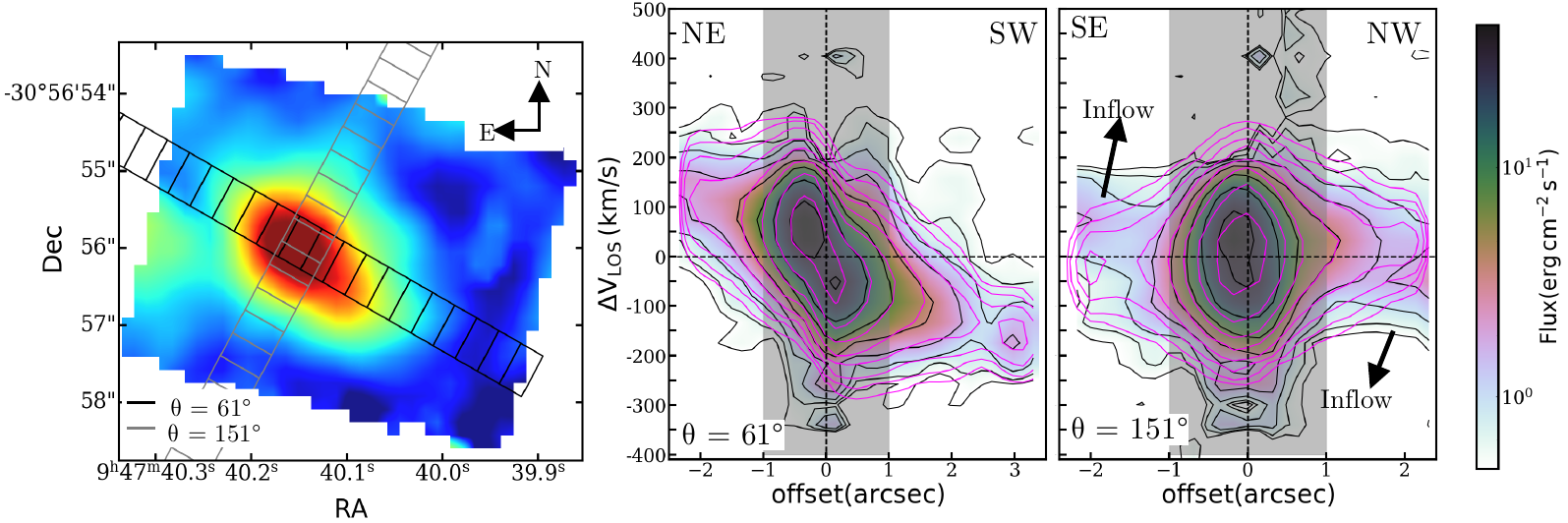}
\includegraphics[width=1.6\columnwidth, trim={5 5 0 0},clip]{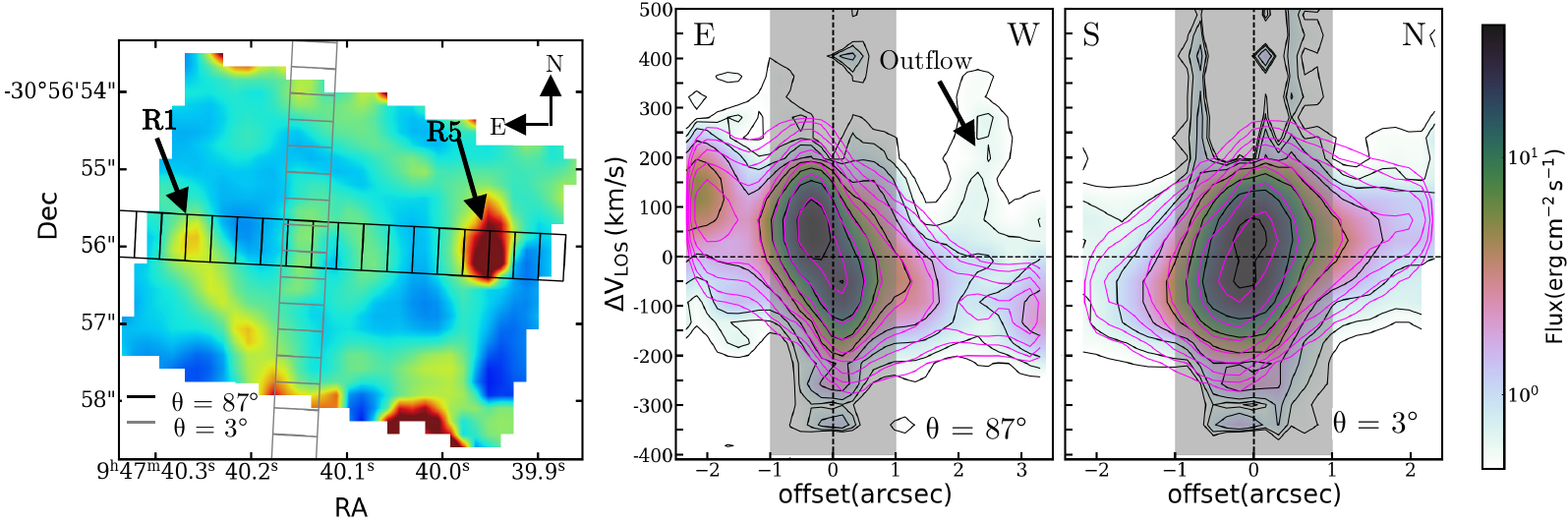} 
\caption{Same as in Fig. \ref{fig:plots_pvs1}, but for S(3). The black arrows show two high velocity dispersion regions (R1 and R5 in Fig. \ref{fig:Maps_with_regions}).}
\label{fig:plots_pvs3}
\end{figure*}

\begin{figure*}
\centering
\includegraphics[width=1.6\columnwidth, trim={5 0 0 0},clip]{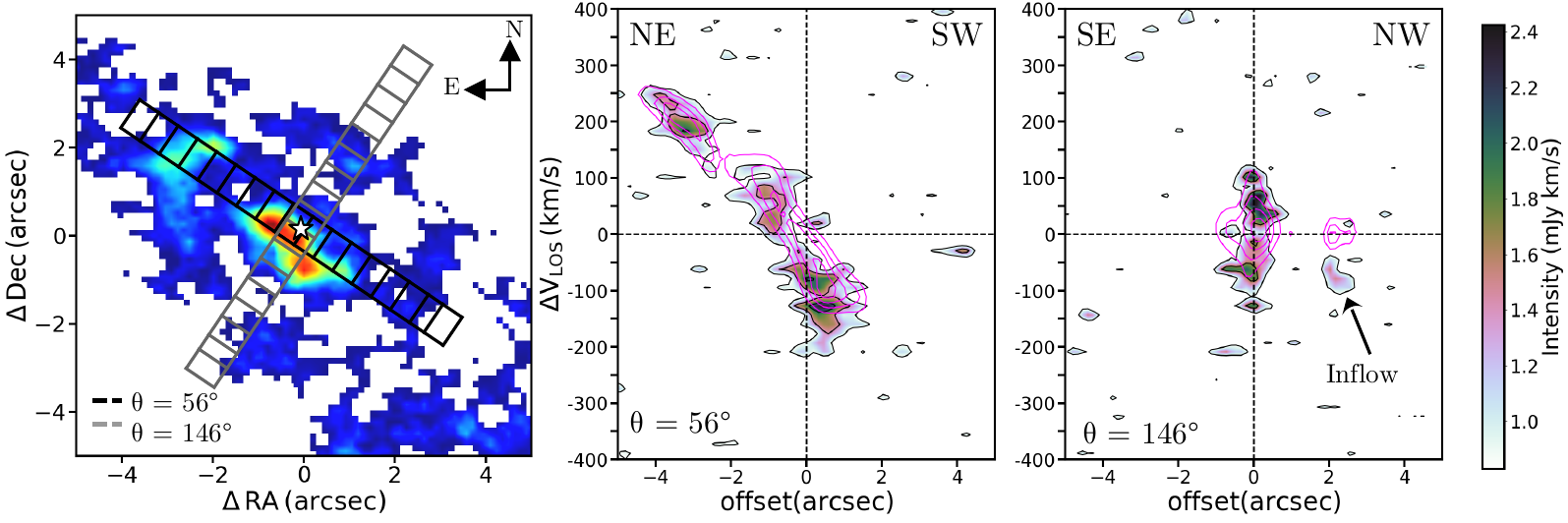} 
\caption{Same as in Fig. \ref{fig:plots_pvs1}, but for CO(2-1). The regions used to extract the PVDs have a size of 0\farcs6~$\times$ 7\arcsec and PAs $= 56\degr$ and 146$\degr$. The contours correspond to 3, 6, 9, 12, 15, 32, and 64 $\sigma$ (= $2 \times 10^{-4}$).} 
\label{fig:plots_pvs_CO}
\end{figure*}

To further characterize and compare the kinematics of the molecular gas from temperatures of tens of Kelvin (CO) to hundreds of Kelvin (H$_2$), we chose three transitions, namely CO(2$-$1), S(1), and S(3). Although S(4) and S(5) would generally be preferable for tracing hotter and more excited molecular gas, we decided not to use them due to the higher number of spaxels with no signal (see Fig. \ref{fig:Maps_H2}). We analyzed the gas kinematics using the ``3D-Based Analysis of Rotating Objects from Line Observations'' (\barolo) software by \cite{DiTeodoro15}. \barolo~allows the user to model rotation in the observed emission line datacubes by fitting a 3D tilted ring. By then subtracting the fitted rotation model from the data, it is possible to investigate the regions that deviate from rotation by inspecting the residuals and position-velocity diagrams (PVDs), as in \citet{Dominguez20} and \citet{Ramos Almeida22}. We used the \textsc{search} parameter in the \textsc{3DFIT} task to find spatial and spectral coherent emissions, applying a mask with a threshold of 3$\sigma_{\rm rms}$. We fixed the central position to the peak of the continuum at 200\, GHz for the CO(2$-$1) emission and to the peak of atomic emission lines lying in the same channels for the H$_2$ lines\footnote{In the case of S(1) we used the position of the peak of [NeIII] and in the case of S(3), of [ArIII].}: at (RA, Dec) of (9:47:40.1472,-30:56:55.922) for S(1) and (9:47:40.1506,-30:56:56.054) for S(3). We allowed the code to fit the observed rotation velocity and velocity dispersion while also varying the PA and the inclination of the disc.

The moment maps of the CO(2$-$1) emission, the corresponding \barolo~model, and the residuals resulting from subtracting the model from the data are shown in Fig. \ref{Fig:MomentMapCO} of Appendix \ref{app:moments}. The moment 0 map shows the mini-spiral and the ring also detected in H$_2$, and the rotation pattern can be reproduced to a certain extent with a disc inclination of i=75$^{\circ}$ and kinematic major axis PA=59$^{\circ}$ (mean values). Significant residuals can be seen in the right panels of Fig. \ref{Fig:MomentMapCO}, indicating important deviations from rotation. We provide an interpretation for these non-circular motions, taking into account the galaxy inclination in Sect. \ref{sec:Discussion}.

In Fig.~\ref{Fig:MomentMapvel}, we show the observed velocity field, rotation model, and corresponding residuals for the three molecular transitions considered here. In the case of the H$_2$ lines, the disc models fitted with \barolo~have inclinations of i=33$^{\circ}$ and 37$^{\circ}$ and major axes PAs of 56$^{\circ}$ and 61$^{\circ}$ for S(1) and S(3). The lower inclination of the H$_2$ disc compared to that derived for the CO is probably related to the smaller FOV of the MIRI/MRS data, which ends in the ring and does not cover the extended and more diffuse emission probed by ALMA on larger scales (see Fig.~\ref{Fig:MomentMapvel}). The rotation pattern is similar for the three transitions considered here, with blueshifted (redshifted) velocities to the southwest (northeast). If we compare the amplitude of the rotation velocities in the same regions, we find similar values among the five H$_2$ lines (see Figs. \ref{fig:Maps_H2} and \ref{Fig:MomentMapvel} and Table \ref{Tab:H2_emission_SameR}). The residuals in the right panels of Fig.~\ref{Fig:MomentMapvel} show high velocities along the minor axis in the outer part of the ring for CO(2$-$1) and S(1), indicating the presence of non-circular motions. In the S(3) velocity map, the high-velocity dispersion knot at $\sim$2.5\arcsec~westward of the nucleus shows ``forbidden'' velocities, that is,  redshifted velocities in the blueshifted rotating side of the galaxy (see bottom panels in Fig. \ref{Fig:MomentMapvel}). The high-velocity residuals in this region, which appear devoid of CO(2-1), highlight how it deviates from regular rotation.

In Fig. \ref{Fig:MomentMapdisp}, we show the velocity dispersion maps, models, and residuals resulting from \barolo~for the three transitions here considered. The warm molecular gas traced by S(1) and S(3) shows higher velocity dispersion than CO in the nucleus ($\sim$130-140 km~s$^{-1}$), and in the case of S(3) also in R5 ($\sim$160 km~s$^{-1}$). The CO(2$-$1) map reveals extended regions of high-velocity dispersion throughout the disc, including areas that are outside the MIRI/MRS FOVs. \barolo~is unable to reproduce these high central dispersion regions, and therefore the residuals are large in the center for the warm molecular gas and across the disc for CO(2$-$1). In the case of the MIRI data, the high nuclear velocity dispersions might be partly due to beam smearing, and we note that BAROLO only reproduces maximum values of 80-90 km/s. The high-velocity dispersion knot in R5 stands out in the residual map of S(3).

Finally, to further investigate the kinematics of the high-velocity dispersion regions, we created PVDs for the S(1) and S(3) emission lines, which are shown in Figs. \ref{fig:plots_pvs1} and \ref{fig:plots_pvs3}. We extracted them using a pseudo-slit width of 0\farcs6~($\sim$100 pc). In the first column of Figs. \ref{fig:plots_pvs1} and \ref{fig:plots_pvs3}, we show the position and orientation of the regions used to produce the PVDs on top of the flux and velocity dispersion maps of each line. The second and third columns of the top row show the PVDs extracted along the major and minor kinematic axes derived from \barolo, and those in the bottom correspond to the PVDs extracted along PAs chosen to capture the regions of interest (i.e., of high-velocity dispersion). We also included the PVDs generated from the model datacube created by \barolo: we produced PVDs extracted in slits with the same width and PAs described above and overlaid them as red contours on the S(1) and S(3) PVDs. For S(1) and S(3), we observe a steep velocity gradient that goes from 300 to $-$250 km~s$^{-1}$ along the kinematic major axis\footnote{We note that PVDs show observed velocities, while the velocity maps shown in Fig. \ref{fig:Maps_H2} correspond to average velocities.}. This velocity gradient is less steep along the minor axis for both lines. The nuclear high-velocity gas that is observed in all the PVDs between $-$1\arcsec~and 1\arcsec~ is an artifact of the continuum subtraction. In the left and central bottom panels of Fig.~\ref{fig:plots_pvs3}, we observe a high-velocity dispersion blob about $\sim$2.5\arcsec~westward of the nucleus, which we labeled R5. This blob is located in a ``forbidden'' quadrant of the rotation pattern and indicates a non-circular gas component that reaches velocities of $\sim$350 km~s$^{-1}$. On the opposite side of the nucleus, at $\sim$1\farcs5~to the east, there is another blob of high-velocity dispersion (labelled as R1 in Fig.~\ref{fig:plots_pvs3}) that can be identified in the same PVD, and that reaches similar velocities. Apart from these two high-velocity and high-velocity dispersion features, the PVDs of both S(1) and S(3) show extended gas at lower velocities that is outside the model contours, for example in the S(1) and S(3) PVDs extracted along the minor axis and along PA=3$^{\circ}$ (see right panels in Figs. \ref{fig:plots_pvs1} and \ref{fig:plots_pvs3}). This emission corresponds to gas in the high-velocity dispersion regions at the north and south edges of the S(1) map.

The PVDs of the CO(2-1) emission are displayed in Fig. \ref{fig:plots_pvs_CO}, extracted along the major and minor kinematic axes derived from \barolo. The width and orientation are similar to those used to extract the PVDs of S(1). We observe that most of the emission along the major axis can be accounted for by circular motions\footnote{Note that PA = 56º is similar to the PA chosen with BAROLO (PA = 59º)}. However, there is nuclear emission showing high blueshifted velocities ($\sim-$200 km$^{-1}$) to the southwest outside the model contours. Along the minor axis (PA=146º), we identify non-circular motions in two regions, one to the northwest ($\sim-$100 km$^{-1}$) and the other to the southeast, near the nucleus ($\sim-$200 km$^{-1}$). These regions correspond to non-circular motions, which we discuss in Sect. \ref{sec:Discussion}.

\subsection{Properties of the warm molecular gas}
\label{diagrams}

\begin{figure*}
\centering
\includegraphics[width=0.65\columnwidth, trim={8 3 20 25},clip]{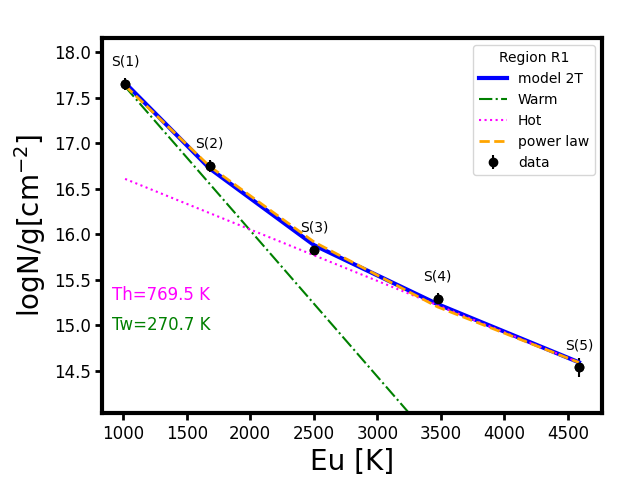}
\includegraphics[width=0.65\columnwidth, trim={8 3 20 25},clip]{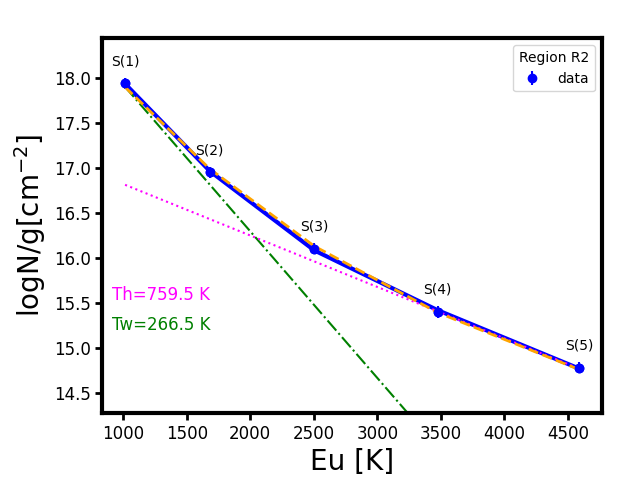}
\includegraphics[width=0.65\columnwidth, trim={8 3 20 25},clip]{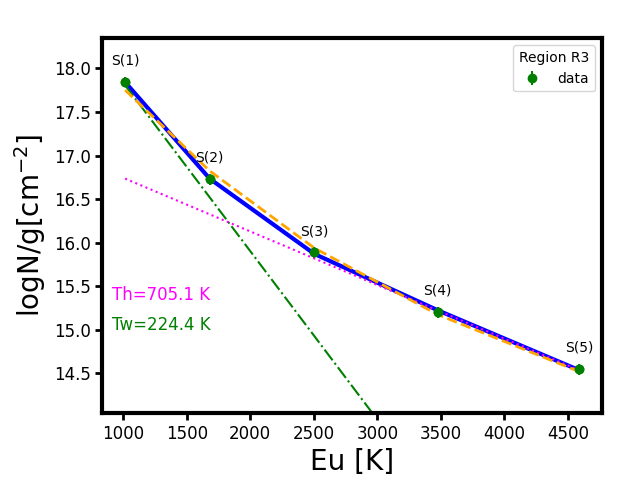}
\includegraphics[width=0.65\columnwidth, trim={8 3 20 25},clip]{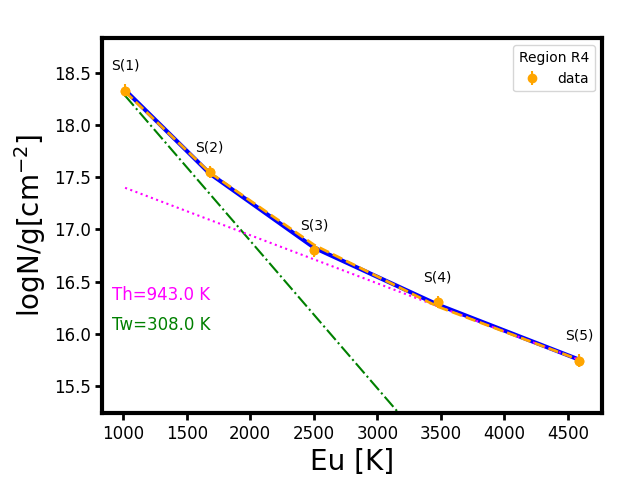}
\includegraphics[width=0.65\columnwidth, trim={8 3 20 25},clip]{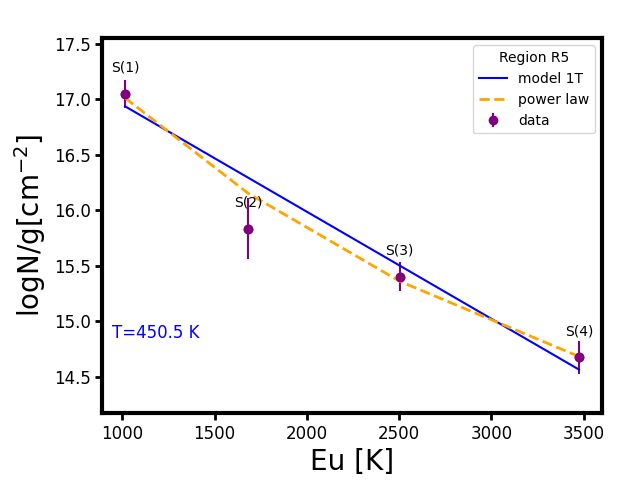}
\includegraphics[width=0.65\columnwidth, trim={8 3 20 25},clip]{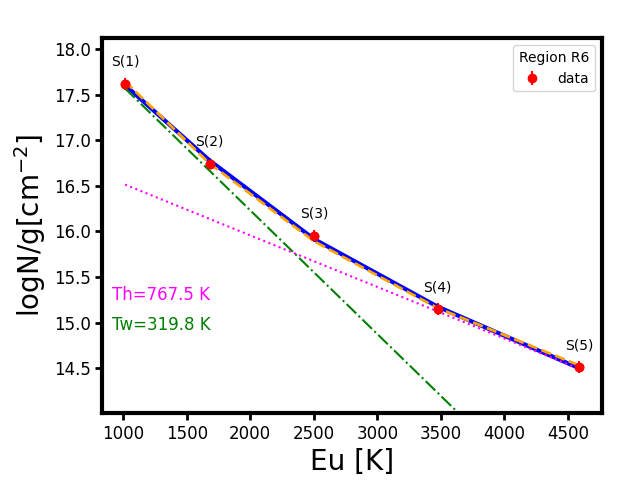}
 \caption{H$_2$ rotational diagrams of the six regions indicated in Fig. \ref{fig:Maps_with_regions}. The dashed yellow and solid blue lines are the fits using a single power law (PL) and a two-temperature model (2T), respectively. The dot-dashed green and dotted magenta lines correspond to the warm and hot components of the 2T model, respectively.}
\label{fig:Temp}
\end{figure*}

We created rotational diagrams following the methodology described in \cite{Rigopoulou02} and \cite{Pereira-Santaella14}. For this, we used the line fluxes derived for the six regions defined in Sect. \ref{subsec:CommonRegions} (see Fig. \ref{fig:Maps_with_regions}), corrected for extinction by adopting the value of A$_V$=1 mag from \citet{Prieto14} and the \citet{Cardelli89} reddening law\footnote{This extinction correction has negligible impact on the line fluxes except for S(3), which is close to the silicate absorption feature.}. Rotational diagrams compare the column density in the upper level of each transition ($N_u$), normalized by its statistical weight $g(u)$ as a function of the temperature $T(u)$ associated with the upper-level energy $E_u$, and they are derived from the Boltzmann equation:

\begin{equation}
    \frac{N_u}{N} = \frac{g(u)}{Z(T_{ex})} \times e^{-\frac{T_u}{T_{ex}}}
\end{equation}

\noindent where $T_{ex}$ is the excitation temperature, $Z(T_{ex})$ is the partition function at $T_{ex}$, and $N_u$ is calculated as:

\begin{equation}
    N_u = \frac{flux(u)}{A(u) \times h \times \nu(u)} \times \frac{4 \pi}{\Omega_{beam}}
\end{equation}

\noindent where $flux(u)$ is the flux of a line in the $u$th state, A$(u)$ is the Einstein A-coefficient of that transition, $\nu(u)$ is the frequency of the transition, $h$ is Planck's constant, and $\Omega_{beam}$ is the area of the region used to extract the flux, calculated as $\Omega_{beam} = \pi r^2$, where $r\, \rm{= 0.5}$\arcsec. We derived the warm ($T < 500\,$K) and hot ($T > 500\,$K) molecular gas masses of the different regions from these excitation diagrams, which use the S(1) to S(5) transitions. The critical densities of these transitions are relatively low, with $n_{H_2} (crit) = 10^2 - 10^5$ cm$\rm{^{-3}}$ at 500 K \citep{LeBourlot99}. Therefore, we can assume local thermodynamic equilibrium (LTE) conditions \citep{Roussel07}. 

We fit the rotational diagram of the six regions using two models: 1) a single power law temperature distribution\footnote{We assumed upper and lower temperatures of 3500 K and 200 K.} (red solid line) and 2) a two-temperatures model, which represents a warm and a hot component (green dashed and dotted lines in Fig. \ref{fig:Temp}). In the case of region R5, we just fitted one temperature component because there is no S(5) measurement there. At first glance, we notice that the excitation temperature increases as the energy level increases, indicating that the gas consists of several components with different temperatures \citep{Rigopoulou02}. In Table \ref{Tab:Outflow_properties}, we report the $\rm{\chi^2/d.o.f.}$ (degrees of freedom), temperatures, column densities, and gas masses for each region. These values are obtained by fitting the S(1) to S(5) transitions using either the single power law or the two-temperatures model. In the two-temperature model, the warm component dominates the S(1) and S(2) emission, and the hotter component dominates the S(3) to S(5) emission. Generally, the models fit the data reasonably well ($\rm{\chi^2/d.o.f. < 1.9}$).

Excluding region R5, the average ($\pm$ the standard deviation of the values reported in Table \ref{Tab:Outflow_properties}) temperatures for the warm and hot components of the two-temperature model are $\rm{T_w = 278\pm38}$ K and $\rm{T_h = 789\pm90}$ K, respectively. The excitation temperatures of the hot component are consistent with those reported in previous works and are typical temperatures in AGN \citep[e.g.][]{Davies05}. The difference between $\rm{T_w}$ and $\rm{T_h}$ is $\sim$500 K, with the warm component being the main contributor to column density and mass. The average column density and mass values of all the regions, excluding R5, are $\rm{\log(N_H^w/cm^{-2})=20.3\pm0.2}$ and (9.6$\pm$5.7)$\times$10$^4$ M$_{\sun}$ for the warm component, and $\rm{\log(N_H^h/cm^{-2}) =18.6\pm0.4}$ and (1.5$\pm$2.9)$\times$10$^4$ M$_{\sun}$ for the hot component. In regions R1, R2, R3, and R6, the mass of the hot component is less than 3\% of the mass of the warm component. In region R4 we find the largest contribution from the hot component, of 1.4$\times$10$^4$ M$_{\sun}$ (see Table \ref{Tab:Outflow_properties}), likely due to its position, closer to the galaxy nucleus than other regions (see Fig. \ref{fig:Maps_with_regions}). In the case of region R5, the knot of high-velocity dispersion in S(3), we measure a temperature of 450 K, a column density of $\rm{\log(N_H/cm^{-2})= 19.0}$, and a gas mass of 4$\times 10^{3}$ M$_{\sun}$, which represents just 4\% of the average warm gas mass obtained for the other regions. This high temperature comes from the fit using only one temperature component, as S(5) measurement is lacking in this region, resulting in high uncertainty for this measurement.

Additionally, in the last two rows of Table \ref{Tab:Outflow_properties}, labeled as Total, we report the values obtained using the H$_2$ fluxes measured in an aperture of 3\farcs6~($\sim$630 pc) in diameter. We found that the difference between the hot and the warm temperatures is $\sim$550 K. The mass of the warm gas is $\rm{\sim 6.4 \times \,  10^5\, M_{\odot}}$, which falls within the range estimated by \citet{Roussel07} for a sample of low-ionization nuclear emission-line region (LINER) and Seyfert galaxies, and the mass of hot gas is $\rm{\sim 2.7 \times \,  10^4\, M_{\odot}}$. We also calculated the mass of cold molecular gas using the CO(2$-$1) flux integrated in the same 3\farcs6~diameter aperture (see dashed white circle in Fig. \ref{fig:MCG-05-23-16_HST}). We used equation (3) of \citet{Solomon05} to compute the CO(2$-$1) line luminosity, $L^\prime_{CO}$, in K km s$^{-1}$pc$\rm^{2}$ units. Then, we assumed that the CO emission is thermalized and optically thick ($R_{21}$=$L^\prime_{CO(2-1)}$/$L^\prime_{CO(1-0)}$=1), and adopted the Milky-Way CO-to-H$_2$ conversion factor \citep[$\alpha_{CO}$=4.36$\pm$1.30 $\rm{(K~km~s^{-1}pc^2)^{-1}}$;][]{Bolatto13} to calculate a mass of $M_{\rm H_2}$=(1.36$\pm$0.50)$\times$10$^7$M$_\odot$. The warm gas mass measured from the rotational H$_2$ lines represents less than 5\% of the total molecular gas estimated from CO(2$-$1). The major source of uncertainty for this mass gas value is the assumed conversion factor, which could be up to a factor of 3$-$4 lower\footnote{If instead of using the Milky-Way CO-to-H$_2$ $\alpha_{CO}$ factor we assume the typical value of ULIRGs of 0.8 (K~km~s$^{-1}$ pc$^2$)$^{-1}$ \citep{Narayanan12}, we measure M$_{\rm H_2}$ = (2.49$\pm$0.5) $\times$ 10$^6$M$_{\odot}$. Assuming this mass, we obtain a warm/cold gas mass ratio equal to 0.25, which also agrees with previous works.}. However, our warm-to-cold gas mass ratio is among the highest reported for nearby ULIRGs by \citet{Higdon06} using data from Spitzer, which are typically less than 1\%. Furthermore, our fraction of 5\% is within the range of values obtained by \citet{Pereira-Santaella14} for a sample of six local infrared bright Seyfert galaxies also observed with Spitzer, of 5--50\%. Recently, \cite{Pereira-Santaella24} calculated the $\alpha_{CO}$ value for NGC\,3256 using various methods with JWST/NIRSpec observations. They derived $\alpha_{CO}$ values for different regions of this galaxy, finding that these values range from 0.26 $-$ 1.09 $\rm{(K~km~s^{-1}pc^2)^{-1}}$. The median $\alpha_{CO}$ values were 0.40 and 0.61 $\rm{(K~km~s^{-1}pc^2)^{-1}}$ using one and two-temperatures models, respectively. These $\alpha_{CO}$ values are 1.3-2 times lower than typically assumed for ULIRGs. Using these median $\alpha_{CO}$ values, we calculated a warm-to-cold gas mass ratio of 0.51 and 0.33.

\begin{table}
    \caption{Parameters derived from the power law (PL) and two-temperatures (2T) fits shown in Fig. \ref{fig:Temp}.}
        \label{Tab:Outflow_properties}
        \centering
        \begin{scriptsize}
        \renewcommand{\arraystretch}{}
        \setlength{\tabcolsep}{0.9pt}
           \begin{tabular}{llcc|ccc|ccc}
            \hline \hline
            Region & Fit & $\chi^{2}/d.o.f.$ & $\beta$ & $\rm{T_w}$ &$\rm{\log (N_{H}^w) }$ &  $\rm{M_w}$ & $\rm{T_h}$ & $\rm{\log (N_{H}^h)} $ & $\rm{M_h}$ \\
            name & model &  &  & (ºK) & ($\rm{cm^{-2}}$) & $\rm{10^5 \, M_{\odot}}$ & (ºK) & ($\rm{cm^{-2}}$) &  $\rm{10^5 \, M_{\odot}}$ \\ 
            \hline
            R1 & PL & 1.28  &  5.03  & $-$ &  20.1  &  0.55  & $-$ & $-$ & $-$ \\
               & 2T & $1.84$ & $-$ & $ 271  \pm  54 $ & $ 20.1 $ & $ 0.50$ & $770  \pm  171 $ & $ 18.5 $ & $ 0.01 $  \\
            \hline
            R2 & PL & 0.36  &  5.23  & - &  20.4  &  1.05  & - & -  & -\\
              & 2T &  0.24  & - & $ 266  \pm  16 $ & $ 20.4 $ & $ 1.0 $ & $ 760  \pm  46 $ & $ 18.7  $ & 0.02  \\
             \hline
            R3 & PL & 1.89  &  5.42  & - &  20.3  &  0.77  & - &- &- \\
              & 2T & 0.22 & - &$224\pm13$ & $20.5$ & $1.41$ & $705 \pm 29$ & $18.7 $ & 0.02 \\
            \hline
            R4 & PL & 0.41  &  4.11  & - &  20.7  &  2.16  & - & - & - \\
               & 2T & 0.42 & - & $308\pm34$ & $20.6$ & 1.62 & $943\pm100$ & $19.0$ & 0.67 \\
            \hline
            R5 & PL & 0.34  &  4.78  & - &  19.5  &  0.12  & - & - & - \\
               & 1T & 1.05 & - & $450\pm55$ & $19.0$ & 0.04 &  - &  - &  -  \\
            \hline
            R6 & PL & 0.44  &  5.18  & - &  20.2  &  0.58  & - & - & - \\
             & 2T &  0.86 & - & $320\pm43$ & $20.0$ & 0.30 & $767\pm144$ & $18.0$ & 0.01  \\ 
            \hline \hline
            Total & PL & 0.64  &  4.28  & - &  20.1  & 6.98 & - & - & -  \\
            & 2T & 0.48  & - & $ 277  \pm  35 $ & $ 20.1 $ & $ 6.38 $ & $ 834  \pm  71 $ & $ 18.7 $ & $ 0.27 $ \\
             \hline
            \end{tabular}
    \end{scriptsize}
    \tablefoot{The results from the fit with the single power law temperature distribution (PL), which represents both the hot and warm gas and assumes that the column density follows a power law ($dN \sim T^{\beta} dT$), are shown in the same columns as those corresponding to the warm component of the two temperatures (2T) fits.}
\end{table}

\begin{figure}
\centering
\includegraphics[width=0.8\columnwidth, trim={25 5 20 20},clip]{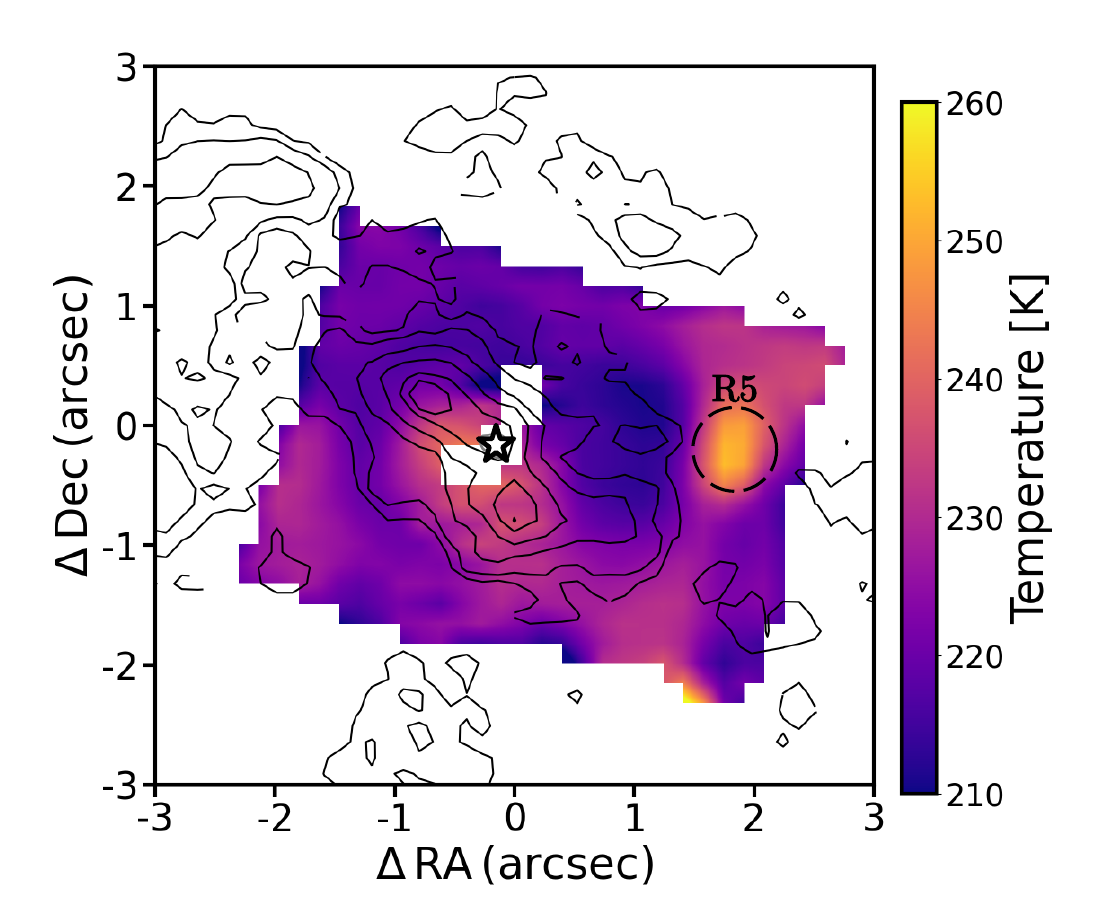}
 \caption{$\rm{H_2}$ excitation map derived from the S(1)/S(3) line ratio (see Sect. \ref{diagrams}). The dashed black circle marks the region R5, also indicated in Fig. \ref{fig:Maps_with_regions} and with an arrow in Fig. \ref{fig:plots_pvs3}. This region has the highest temperature area on this map, with a temperature $\sim$260 K. The black solid contours correspond to 0.2, 0.4, 0.6, 0.8, and 1$\sigma$ of the CO(2-1) emission. The white star shows the AGN location according to the ALMA continuum.}
\label{fig:TempS1S3}
\end{figure}

Finally, in Fig. \ref{fig:TempS1S3} we show an excitation map of MCG$-$05$-$16$-$23 using the S(1)/S(3) line ratio in the FOV of Ch2-m, determined by the S(3) line (see Table \ref{tab:fwhm}) and following the same procedure as in \cite{Davies24}. The nuclear spiral and the ring show higher temperatures than the inter-arm gas, and the S(3) high-velocity dispersion knot (R5 in Fig. \ref{fig:Maps_with_regions}) shows the highest temperature of $\sim$260 K as determined from the S(1)/S(3) line ratio. In the next section, we will discuss the possible origin of this region.

\section{Discussion}
\label{sec:Discussion}

As it can be seen from Fig. \ref{fig:MCG-05-23-16_HST}, MCG$-$05$-$23$-$16 shows a dust lane in the optical HST/WFC3 image that resembles a nuclear-trailing spiral and a larger-scale ring (see also Fig. \ref{Fig:ALMA_vs_HST} in Appendix \ref{app:ALMAvsHST}). The dust lane, of $\sim$2\arcsec~extent ($\sim$360 pc) and oriented along the major axis of the galaxy, was previously reported by \citet{Prieto14}. The CO(2$-$1) and H$_2$ moment maps reveal that the nuclear spiral is connected with the ring, of major axis $\sim$8\arcsec~(1.4 kpc), through a network of arms/filaments (see Figs. \ref{Fig:MomentMapCO} and \ref{fig:Maps_H2}). A nuclear trailing spiral connected with a ring is indicative of gas that is losing angular momentum and feeding the SMBH in the presence of a bar (e.g., \citealt{Regan03, Buta13, Shimizu19, Mondal21}). According to several studies, circumnuclear rings are formed near the inner Lindblad resonances \citep[ILR,][]{Lindblad64} due to the bar potential when the gas loses angular momentum and accumulates in that region (see references in  \citealt{Schmidt19}). Recently, \cite{Sormani24} proposed that a nuclear ring is actually the inner edge of an extensive gap that opens around the ILR due to the excitation of waves by a bar potential.

Inside the nuclear ring at the ILR, the gravity torques due to the presence of a nuclear bar are negative and induce the formation of a trailing nuclear spiral \citep[e.g.][]{Kim17}, favoring the transport of gas towards the center, as seen in other observations of nearby Seyfert galaxies \citep[see][]{Combes14, Audibert19, Audibert21}. The winding pattern of the nuclear spiral is predicted due to the presence of a massive black hole in the center \citep{Buta96}, and in the case of MCG$-$05$-$23$-$16, it seems to coincide with the trailing spiral arms at large scale, according to the reported geometry of the galaxy disk (i.e., the northwest side is the far side), and the observed sense of rotation of the gas. The computation of the gravity torques due to the nuclear bar is beyond the scope of this paper. However, the presence of a nuclear bar in MCG$-$05$-$23$-$16 is further supported by the S-shape zero isovelocities observed in the CO(2-1) velocity map shown in Fig. \ref{Fig:MomentMapCO}, which are not aligned with the minor axis \citep[see e.g.][for references on S-shape velocities as bar tracers]{Sanders80, Pence84, Randriamampandry15, Cohen20}.

On the largest scales that can be probed with S(1), which coincide with the ring, we see evidence of strong non-circular motions along the minor axis both in CO and H$_2$. Considering that the southeast is the near side of the galaxy, based on both the higher extinction there (see right panel of Fig. \ref{fig:MCG-05-23-16_HST} and \citealt{Prieto14}) the velocity residuals along the kinematic minor axis (right panels of Fig. \ref{Fig:MomentMapvel}) would correspond to the radial projection of local inflowing gas motions in the plane of the galaxy (positive velocities in the southeast and negative velocities in the northwest). This is also seen in the PVD diagram of S(1) along the minor axis (see top right panel of Fig. \ref{fig:plots_pvs1}): there is gas within the ring that is not reproduced by the rotating disc model fitted with \barolo~that is blueshifted to the northwest and redshifted to the southeast. Something similar is seen in S(3), but to a lesser extent due to the smaller FOV of Ch2. These residuals along the minor axis are relatively far from the nuclear spiral, and they can be better explained by the radial projection of non-circular motions typical of the elliptical gas orbits in the presence of a bar, rather than by being the signature of a truly radial inflow.

Evidence of non-circular motions that could be consistent with outflowing gas, although not aligned with the kinematic minor axis, are found from the S(3) kinematics, and to a lesser extent, from S(1). The velocity maps of these emission lines show red residuals at $\sim$2.5\arcsec~(440 pc) westward of the nucleus (region R5 in Fig. \ref{fig:Maps_with_regions}, at PA$\sim$270\degr; see last column in Fig. \ref{Fig:MomentMapvel}) and in the case of S(3), also velocity dispersion residuals (see Fig. \ref{Fig:MomentMapdisp}). These residuals are consistent, in principle, with outflowing gas, and they can be seen in the PVD along PA=87\degr~shown in Fig. \ref{fig:plots_pvs3}, clearly deviating from the rotation disc model fitted with \barolo.

This area of higher velocity dispersion in S(3) coincides with an area devoid of CO(2$-$1) emission (see Figs. \ref{fig:Maps_H2} and \ref{fig:Maps_with_regions}). This suggests that the more turbulent, hotter, and excited gas detected with S(3) fills regions depleted of cold gas. This is observed when we compare the morphology of CO and those of the different H$_2$ lines observed by MIRI (see Fig. \ref{fig:Maps_H2}): the CO traces the nuclear spiral, the ring, and the connecting arms, and the regions that are devoid of CO are occupied by warmer molecular gas showing the highest values of the velocity dispersion there. Considering the compact and localized morphology of the R5 knot seen in S(3), it is possible that that gas might have been accelerated or disrupted by star formation there. This R5 knot cannot be explained by non-circular motions in the plane of the galaxy because they are well outside the expected virial range, and they show an opposite sign of the velocities (see middle panel of Fig. \ref{fig:plots_pvs3}). A jet-driven outflow could be another possible scenario for explaining the kinematics, but we do not see any spatial correspondence with the jet (black contours in Fig. \ref{fig:Maps_with_regions}). A localized, star-formation-driven outflow is supported by the presence of Polycyclic Aromatic Hydrocarbon (PAH) emission in the area between the nucleus and the S(3) knot, as traced by the 11.3 $\mu$m feature. This PAH band has been used as a reliable tracer of recent star formation in AGN \citep[e.g.][]{Diamond-Stanic10, Diamond-Stanic12, Esparza-Arredondo18}, which is also found to be the most resilient PAH band in the harsh environment of AGN and their outflows \citep{Garcia-Bernete22, Garcia-Bernete24b}.

\begin{figure}
\centering
\includegraphics[width=1.\columnwidth, trim={10 0 0 15},clip]{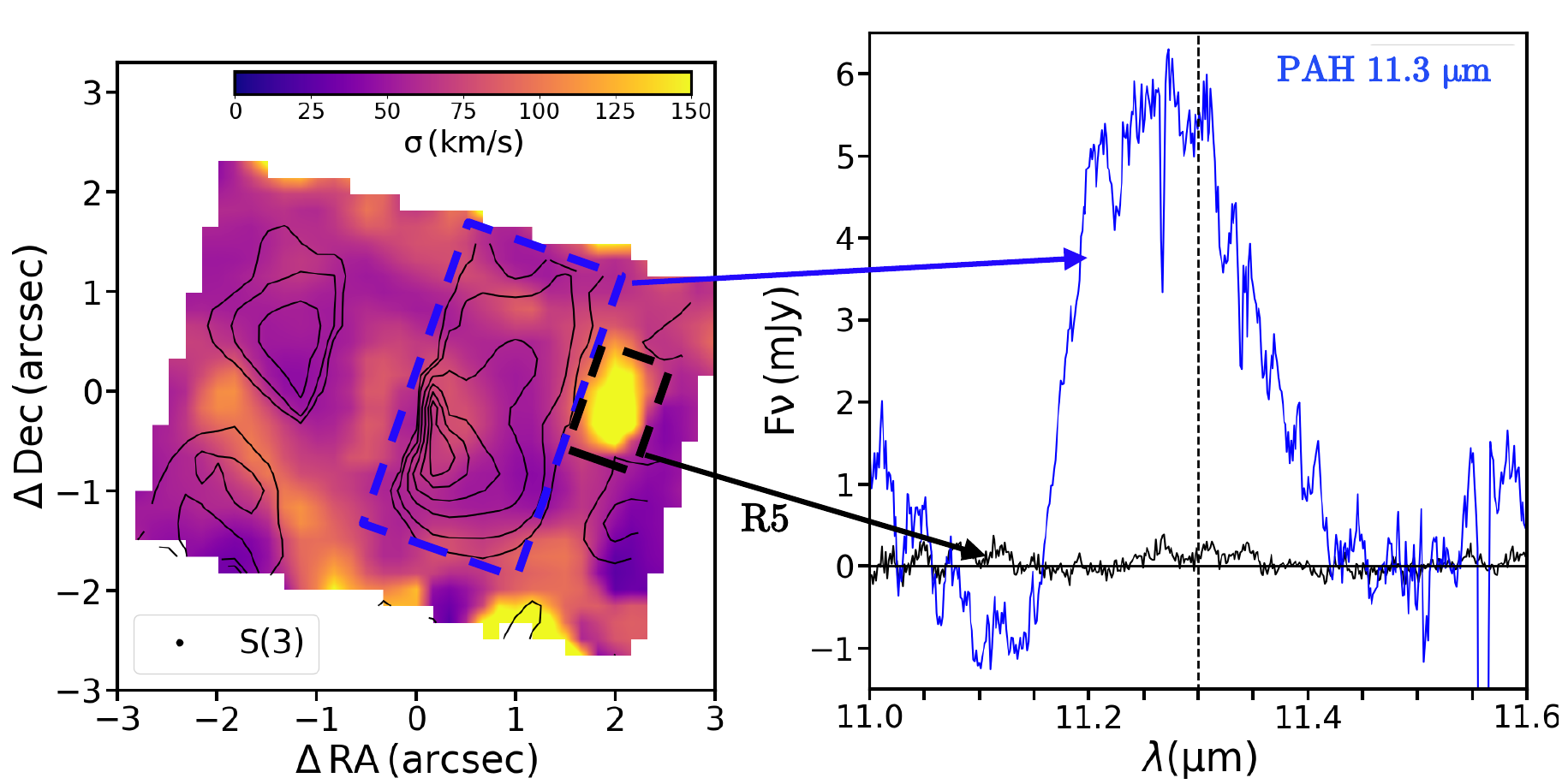}
 \caption{The left panel shows the S(3) velocity dispersion map with contours of the 11.3 $\mu$m PAH feature overlaid in black. The right panel shows the profiles of the 11.3 $\mu$m PAH feature extracted from the regions indicated in the left panel.}
\label{fig:PAH113}
\end{figure}

Using the integrated luminosity of the PAH feature extracted from the region indicated with a blue dashed rectangle in Fig. \ref{fig:PAH113} ($L_{\rm 11.3~\mu m}$ = 7.34 $\times 10^{39}$ erg~s$^{-1}$)\footnote{$L_{\rm 11.3~\mu m}$ was multiplied by a factor of two to make the SFR comparable with PAHFIT-derived measurements.}  and Eq.~12 in \citet{Shipley16}, as in \citet{Ramos Almeida23}, we measured a star formation rate (SFR) of 0.013$\pm^{0.005}_{0.004}$ M$_{\odot}~yr^{-1}$. Gas compression induced by the inflowing gas might have triggered star formation locally, and the young stars producing the PAH emission there could have driven a localized warm molecular outflow in the region shown with a black dashed rectangle in Fig. \ref{fig:PAH113}. The mass of the H$_2$ knot (R5) calculated in Sect. \ref{diagrams} is $\rm{4000~M_{\odot}}$ (see Table \ref{Tab:Outflow_properties}, 1T model), but part of the gas there is low-velocity gas participating in rotation (negative velocities at $\sim$2.5\arcsec~to the west in the central bottom panel of Fig. \ref{fig:plots_pvs3}) and part is outflowing (positive velocities at the same position). Therefore, the outflow mass will be lower than the mass estimated for R5 as a whole, so we consider the latter as an upper limit to the outflow mass. 
By then using a velocity representative of the outflowing gas from the PVD shown in Fig. \ref{fig:plots_pvs3}, of $\sim$200 km~s$^{-1}$, and a radius of $\sim$0\farcs5~(88 pc, which is the R5 radius), we derive a mass outflow rate of \.M$_{out}=M_{H2}^{R5} \times v_{ out} \times r_{out}^{-1}<0.01$ $~M_{\sun}~yr^{-1}$ and kinetic energy of <$\rm{1.2 \times 10^{38}}$ erg~s$^{-1}$ \citep[see Eq. 3 in][]{Speranza24}. According to the estimated values of the SFR and mass outflow rate, star formation might be enough to drive the outflow (mass loading factor = \.M$_{\rm out}$/SFR < 0.8), although the uncertainties are significant.

As mentioned before, despite the presence of extended radio emission that is consistent with a compact jet of $\sim$200 pc in size and PA$\sim -$10\degr~\citep{Orienti10}, we do not find evidence of a strong jet-ISM interaction in MCG$-$05$-$23$-$16, at least in the case of the molecular gas. If the radio emission corresponds to a jet, we would be seeing the approaching side, considering its morphology: elongated and brighter towards the north. In that case, and considering that the northwest is the far side of the galaxy, the jet would be subtending a large angle relative to the CO and H$_2$ discs, and therefore, it will have a small impact on the ISM \citep{Meenakshi22,Ramos Almeida22,Audibert23}.

\section{Conclusions}
\label{sec:Conclusions}

In this work, we characterize for the first time the physical properties (temperature, column density, and mass), morphology, and kinematics of the molecular gas in the spheroidal galaxy MCG$-$05$-$23$-$16, using data from the MRS module of JWST/MIRI and ALMA. Thanks to these datasets, we studied the connection between the warm ($ T \sim$100$-$1000 K) and the cold ($T<100$ K) molecular gas traced by the rotational mid-infrared H$_2$ lines and the sub-mm CO(2$-$1) transition.

Our key findings include the detection of a nuclear trailing spiral of $\sim$350 pc in size and a surrounding ring of $\sim$1.8 kpc in both CO(2$-$1) and $\rm{H_2}$, which are also identified through optical HST observations. These features indicate the typical gas response to a stellar bar, where the ring may correspond to the ILR of the bar and the trailing nuclear spiral inside the ring is triggering an ongoing fueling episode of the AGN.

The morphology and kinematics of the gas traced with the $\rm{H_2}$0$-$0S(1), S(2), and CO(2$-$1) lines are similar, with rotational velocities of up to $\pm$250 km~s$^{-1}$ and non-circular motions on the spatial scales of the ring that are consistent with the projection of non-circular motions associated with the elliptical gas orbits followed by the gas. On the other hand, the warmer and more excited molecular gas traced with the S(3), S(4), and S(5) lines shows more complex kinematics that include clumps of high-velocity dispersion (of up to $\sim$160 km~s$^{-1}$) in regions where CO(2$-$1) is undetected. One of these clumps, located at $\sim$350 pc westward of the nucleus, shows kinematics that is consistent with outflowing gas, which could be driven by recent star formation in adjacent regions, traced by the 11.3 $\mu$m PAH feature. The compact jet previously detected through VLA observations does not appear to impact the molecular gas, probably because its orientation is not favorable for it to entrain the molecular gas disc.  

Overall, we observe a stratification of the molecular gas. The colder gas is located in the nuclear spiral, ring, and connecting arms, while most warmer gas with higher velocity-dispersion fills the space between them. This work highlights the potential synergy between JWST and ALMA observations in advancing the study of molecular gas in galaxies.

\begin{acknowledgements}
We thank the referee for their useful comments and suggestions. D.E.A., C.R.A., and A.A. acknowledge support from the Agencia Estatal de Investigaci\'on of the Ministerio de Ciencia, Innovaci\'on y Universidades (MCIU/AEI) under the grant ``Tracking active galactic nuclei feedback from parsec to kiloparsec scales'', with reference PID2022$-$141105NB$-$I00 and the European Regional Development Fund (ERDF).
D.E.A. and B.G.L. acknowledge support from the Spanish Ministry of Science and Innovation through the Spanish State Research Agency (AEI-MCINN/10.13039/501100011033) through grants ``Participation of the Instituto de Astrof\'isica de Canarias in the development of HARMONI: D1 and Delta-D1 phases with references PID2019$-$107010GB100 and PID2022-140483NB-C21 and the Severo Ochoa Program 2020$-$2023 (CEX2019$-$000920$-$S)''.
M.P.S. acknowledges support from grants RYC2021$-$033094$-$I and CNS2023$-$145506 funded by MCIN/AEI/10.13039/501100011033 and the European Union NextGenerationEU/PRTR.
L.H.M. and A.A.H. acknowledge financial support by the grant PID2021$-$124665NB$-$I00 funded by the Spanish Ministry of Science and Innovation and the State Agency of Research MCIN/AEI/10.13039/501100011033 PID2021$-$124665NB$-$I00 and ERDF A way of making Europe.
A.J.B. acknowledges funding from the “FirstGalaxies” Advanced Grant from the European Research Council (ERC) under the European Union’s Horizon 2020 research and innovation program (Grant agreement No. 789056).
E.K.S.H., L.Zhang, C.Packhman acknowledges grant support from the Space Telescope Science Institute (ID: JWST$-$GO$-$01670.007$-$A).
S.G.B. acknowledges support from the Spanish grant PID2022-138560NB-I00, funded by MCIN/AEI/10.13039/501100011033/FEDER, EU. CR acknowledges support from Fondecyt Regular grant 1230345, ANID BASAL project FB210003, and the China-Chile joint research fund. I.G.B. is supported by the Programa Atracci\'on de Talento Investigador ``C\'esar Nombela'' via grant 2023-T1/TEC-29030 funded by the Community of Madrid. E.B.  acknowledges support from the Spanish grants PID2022-138621NB-I00 and PID2021-123417OB-I00, funded by MCIN/AEI/10.13039/501100011033/FEDER, EU. T.D.S. acknowledges the research project was supported by the Hellenic Foundation for Research and Innovation (HFRI) under the "2nd Call for HFRI Research Projects to support Faculty Members \& Researchers" (Project Number: 03382). This work is based on observations made with the NASA/ESA/CSA James Webb Space Telescope. The data were obtained from the Mikulski Archive for Space Telescopes at the Space Telescope Science Institute, operated by the Association of Universities for Research in Astronomy, Inc., under NASA contract NAS 5-03127 for JWST. The specific observations analyzed here can be accessed via \href{https://mast.stsci.edu/portal/Mashup/Clients/Mast/Portal.html?searchQuery=%7B%22service%22:%22DOIOBS%22,%22inputText%22:%2210.17909/vre3-m991%22%7D}{doi:10.17909/vre3-m991}. This paper makes use of the following ALMA data: ADS/JAO.ALMA\#
2019.1.01742.S. ALMA is a partnership of ESO (representing its member states), NSF (USA) and NINS (Japan), together with NRC(Canada) and NSC and ASIAA (Taiwan), in cooperation with the Republic of Chile. The Joint ALMA Observatory is operated by ESO, AUI/NRAO and NAOJ. The National Radio Astronomy Observatory is a facility of the National Science Foundation operated under a cooperative agreement by Associated Universities, Inc. We used observations made with the NASA/ESA Hubble Space Telescope, and obtained from the Hubble Legacy Archive, which is a collaboration between the Space Telescope Science Institute (STScI/NASA), the Space Telescope European Coordinating Facility (ST-ECF/ESA), and the Canadian Astronomy Data center (CADC/NRC/CSA).

\end{acknowledgements}

%

\begin{thebibliography}{}

    \bibitem[Alonso-Herrero et al.(2021)]{Alonso-Herrero21} Alonso-Herrero, A., Garc{\'\i}a-Burillo, S., H{\"o}nig, S.~F., et al.\ 2021, \aap, 652, A99.

    \bibitem[{\'A}lvarez-M{\'a}rquez et al.(2023)]{Alvarez-Marquez23} {\'A}lvarez-M{\'a}rquez, J., Labiano, A., Guillard, P., et al.\ 2023, \aap, 672, A108.

    \bibitem[Argyriou et al.(2023)]{Argyriou23} Argyriou, I., Glasse, A., Law, D.~R., et al.\ 2023, \aap, 675, A111.

    \bibitem[Asmus et al.(2015)]{Asmus15} Asmus, D., Gandhi, P., H{\"o}nig, S.~F., et al.\ 2015, \mnras, 454, 766.

    \bibitem[Audibert et al.(2019)]{Audibert19} Audibert, A., Combes, F., Garc{\'\i}a-Burillo, S., et al.\ 2019, \aap, 632, A33.

    \bibitem[Audibert et al.(2021)]{Audibert21} Audibert, A., Combes, F., Garc{\'\i}a-Burillo, S., et al.\ 2021, \aap, 656, A60.

    \bibitem[Audibert et al.(2023)]{Audibert23} Audibert, A., Ramos Almeida, C., Garc{\'\i}a-Burillo, S., et al.\ 2023, \aap, 671, L12.

    \bibitem[Bianchin et al.(2024)]{Bianchin24} Bianchin, M., U, V., Song, Y., et al.\ 2024, \apj, 965, 103.

    \bibitem[Bolatto et al.(2013)]{Bolatto13} Bolatto, A. D., Wolfire, M., \& Leroy, A. K. 2013, ARA\&A, 51, 207.

    \bibitem[Briggs(1995)]{Briggs95} Briggs, D.~S.\ 1995, Ph.D. Thesis, New Mexico Institute of Mining and Technology

    \bibitem[Buta \& Combes(1996)]{Buta96} Buta, R. \& Combes, F.\ 1996, \fcp, 17, 95

    \bibitem[Buta(2013)]{Buta13} Buta, R.~J.\ 2013, Secular Evolution of Galaxies, 155. Lecture 2: Barred and spiral galaxies.

    \bibitem[Bushouse et al.(2024)]{Bushouse24} Bushouse, H., Eisenhamer, J., Dencheva, N., et al.\ 2023, Zenodo. JWST Calibration Pipeline (1.10.2). Zenodo. https://doi.org/10.5281/zenodo.7829329
    
    \bibitem[Cardelli et al.(1989)]{Cardelli89} Cardelli, J.~A., Clayton, G.~C., \& Mathis, J.~S.\ 1989, \apj, 345, 245.

    \bibitem[Cazaux \& Tielens(2002)]{Cazaux02} Cazaux, S. \& Tielens, A.~G.~G.~M.\ 2002, \apjl, 575, L29.

    \bibitem[Cohen et al.(2020)]{Cohen20} Cohen, D.~P., Turner, J.~L., \& Consiglio, S.~M.\ 2020, \mnras, 493, 627.
    
    \bibitem[Combes(2003)]{Combes03} Combes, F.\ 2003, Active Galactic Nuclei: From Central Engine to Host Galaxy, 290, 411.

    \bibitem[Combes et al.(2014)]{Combes14} Combes, F., Garc{\'\i}a-Burillo, S., Casasola, V., et al.\ 2014, \aap, 565, A97.

    \bibitem[Dale \& Helou(2002)]{Dale02} Dale, D.~A. \& Helou, G.\ 2002, \apj, 576, 159.

    \bibitem[Dale et al.(2005)]{Dale05} Dale, D.~A., Sheth, K., Helou, G., et al.\ 2005, \aj, 129, 2197.

    \bibitem[Davies et al.(2004)]{Davies04} Davies, R.~I., Tacconi, L.~J., \& Genzel, R.\ 2004, \apj, 602, 148.
    
    \bibitem[Davies et al.(2005)]{Davies05} Davies, R.~I., Sternberg, A., Lehnert, M.~D., et al.\ 2005, \apj, 633, 105.

    \bibitem[Davies et al.(2006)]{Davies06} Davies, R.~I., Thomas, J., Genzel, R., et al.\ 2006, \apj, 646, 754.

    \bibitem[Davies et al.(2009)]{Davies09} Davies, R.~I., Maciejewski, W., Hicks, E.~K.~S., et al.\ 2009, \apj, 702, 114.
    
    \bibitem[Davies et al.(2020)]{Davies20} Davies, R., Baron, D., Shimizu, T., et al.\ 2020, \mnras, 498, 4150.

    \bibitem[Davies et al.(2024)]{Davies24} Davies, R., Shimizu, T., Pereira-Santaella, M., et al.\ 2024, \aap, 689, A263.

    \bibitem[Diamond-Stanic \& Rieke(2010)]{Diamond-Stanic10} Diamond-Stanic, A.~M. \& Rieke, G.~H.\ 2010, \apj, 724, 140.

    \bibitem[Diamond-Stanic et al.(2012)]{Diamond-Stanic12} Diamond-Stanic A. M., Rieke G. H.\ 2012, \apj, 746, 168.
    
    \bibitem[Di Teodoro \& Fraternali(2015)]{DiTeodoro15} Di Teodoro, E.~M. \& Fraternali, F.\ 2015, \mnras, 451, 3021.

    \bibitem[Dom{\'\i}nguez-Fern{\'a}ndez et al.(2020)]{Dominguez20} Dom{\'\i}nguez-Fern{\'a}ndez, A.~J., Alonso-Herrero, A., Garc{\'\i}a-Burillo, S., et al.\ 2020, \aap, 643, A127.

    \bibitem[Dubois et al.(2016)]{Dubois16} Dubois, Y., Peirani, S., Pichon, C., et al.\ 2016, \mnras, 463, 3948.

    \bibitem[Esparza-Arredondo et al.(2018)]{Esparza-Arredondo18} Esparza-Arredondo, D., Gonz{\'a}lez-Mart{\'\i}n, O., Dultzin, D., et al.\ 2018, \apj, 859, 124.

    \bibitem[Ferrarese \& Merritt(2000)]{Ferrarese00} Ferrarese, L. \& Merritt, D.\ 2000, \apjl, 539, L9.
    
    \bibitem[Ferruit et al.(2000)]{Ferruit00} Ferruit, P., Wilson, A.~S., \& Mulchaey, J.\ 2000, \apjs, 128, 139.

    \bibitem[Garc{\'\i}a-Bernete et al.(2022)]{Garcia-Bernete22} Garc{\'\i}a-Bernete, I., Rigopoulou, D., Alonso-Herrero, A., et al.\ 2022, \aap, 666, L5.

    \bibitem[Garc{\'\i}a-Bernete et al.(2024a)]{Garcia-Bernete24a} Garc{\'\i}a-Bernete, I., Alonso-Herrero, A., Rigopoulou, D., et al.\ 2024, \aap, 681, L7.

    \bibitem[Garc{\'\i}a-Bernete et al.(2024)]{Garcia-Bernete24b} Garc{\'\i}a-Bernete, I., Rigopoulou, D., Donnan, F.~R., et al.\ 2024, \aap, 691, A162.
    
    \bibitem[Garc{\'\i}a-Burillo et al.(2005)]{Garcia-Burillo05} Garc{\'\i}a-Burillo, S., Combes, F., Schinnerer, E., et al.\ 2005, \aap, 441, 1011.

    \bibitem[Garc{\'\i}a-Burillo \& Combes(2012)]{Garcia-Burillo12} Garc{\'\i}a-Burillo, S. \& Combes, F.\ 2012, Journal of Physics Conference Series, 372, 012050.
    
    \bibitem[Garc{\'\i}a-Burillo et al.(2021)]{Garcia-Burillo21} Garc{\'\i}a-Burillo, S., Alonso-Herrero, A., Ramos Almeida, C., et al.\ 2021, \aap, 652, A98.

    \bibitem[Garc{\'\i}a-Burillo et al.(2024)]{Garcia-Burillo24} Garc{\'\i}a-Burillo, S., Hicks, E.~K.~S., Alonso-Herrero, A., et al.\ 2024, \aap, 689, A347.

    \bibitem[Gardner et al.(2023)]{Gardner23} Gardner, J.~P., Mather, J.~C., Abbott, R., et al.\ 2023, \pasp, 135, 068001.

    \bibitem[Goodrich et al.(1994)]{Goodrich94} Goodrich, R.~W., Veilleux, S., \& Hill, G.~J.\ 1994, \apj, 422, 521.
    
    \bibitem[Habart et al.(2005)]{Habart05} Habart, E., Walmsley, M., Verstraete, L., et al.\ 2005, \ssr, 119, 71.
    
    \bibitem[Harrison \& Ramos Almeida(2024)]{Harrison24} Harrison, C.~M. \& Ramos Almeida, C.\ 2024, Galaxies, 12, 17.

    \bibitem[Hermosa Mu{\~n}oz et al.(2024)]{Hermosa-Munoz24} Hermosa Mu{\~n}oz, L., Alonso-Herrero, A., Pereira-Santaella, M., et al.\ 2024, \aap, 690, A350.
    
    \bibitem[Herrera-Camus et al.(2020)]{Herrera-Camus20} Herrera-Camus, R., Janssen, A., Sturm, E., et al.\ 2020, \aap, 635, A47.

    \bibitem[Hickox et al.(2014)]{Hickox14} Hickox, R.~C., Mullaney, J.~R., Alexander, D.~M., et al.\ 2014, \apj, 782, 9.

    \bibitem[Hicks et al.(2009)]{Hicks09} Hicks, E.~K.~S., Davies, R.~I., Malkan, M.~A., et al.\ 2009, \apj, 696, 448.

    \bibitem[Higdon et al.(2006)]{Higdon06} Higdon, S.~J.~U., Armus, L., Higdon, J.~L., et al.\ 2006, \apj, 648, 323.

    \bibitem[Ineson et al.(2015)]{Ineson15} Ineson, J., Croston, J.~H., Hardcastle, M.~J., et al.\ 2015, \mnras, 453, 2682.

    \bibitem[Kawamuro et al.(2016)]{Kawamuro16} Kawamuro, T., Ueda, Y., Tazaki, F., et al.\ 2016, \apjs, 225, 14.

    \bibitem[Kim \& Elmegreen(2017)]{Kim17} Kim, W.-T. \& Elmegreen, B.~G.\ 2017, \apjl, 841, L4.

    \bibitem[Kolcu et al.(2023)]{Kolcu23} Kolcu, T., Maciejewski, W., Gadotti, D.~A., et al.\ 2023, \mnras, 524, 207.

    \bibitem[Kormendy \& Richstone(1995)]{Kormendy95} Kormendy, J. \& Richstone, D.\ 1995, \araa, 33, 581. 

    \bibitem[Labiano et al.(2016)]{Labiano16} Labiano, A., Azzollini, R., Bailey, J., et al.\ 2016, \procspie, 9910, 99102W. doi:10.1117/12.2232554

    \bibitem[Labiano et al.(2021)]{Labiano21} Labiano, A., Argyriou, I., {\'A}lvarez-M{\'a}rquez, J., et al.\ 2021, \aap, 656, A57.

    \bibitem[Le Bourlot et al.(1999)]{LeBourlot99} Le Bourlot, J., Pineau des For{\^e}ts, G., \& Flower, D.~R.\ 1999, \mnras, 305, 802.

    \bibitem[Lebouteiller et al.(2011)]{Lebouteiller11} Lebouteiller, V., Barry, D.~J., Spoon, H.~W.~W., et al.\ 2011, \apjs, 196, 8.

    \bibitem[Lindblad(1964)]{Lindblad64} Lindblad, B.\ 1964, Astrophysica Norvegica, 9, 103
    
    \bibitem[Liu et al.(2024)]{Liu24} Liu, V., Zoghbi, A., \& Miller, J.~M.\ 2024, \apj, 963, 38.

    \bibitem[Magorrian et al.(1998)]{Magorrian98} Magorrian, J., Tremaine, S., Richstone, D., et al.\ 1998, \aj, 115, 2285.

    \bibitem[Martini(2004)]{Martini04} Martini, P.\ 2004, The Interplay Among Black Holes, Stars and ISM in Galactic Nuclei, 222, 235.

    \bibitem[Marinucci et al.(2022)]{Marinucci22} Marinucci, A., Muleri, F., Dovciak, M., et al.\ 2022, \mnras, 516, 5907.

    \bibitem[McMullin et al.(2007)]{McMullin07} McMullin, J.~P., Waters, B., Schiebel, D., et al.\ 2007, Astronomical Data Analysis Software and Systems XVI, 376, 127

    \bibitem[Meenakshi et al.(2022)]{Meenakshi22} Meenakshi, M., Mukherjee, D., Wagner, A.~Y., et al.\ 2022, \mnras, 516, 766.

    \bibitem[Mondal \& Chattopadhyay(2021)]{Mondal21} Mondal, D. \& Chattopadhyay, T.\ 2021, Celestial Mechanics and Dynamical Astronomy, 133, 43.

    \bibitem[Mundell et al.(2009)]{Mundell09} Mundell, C.~G., Ferruit, P., Nagar, N., et al.\ 2009, \apj, 703, 802.

    \bibitem[Narayanan et al.(2012)]{Narayanan12} Narayanan, D., Krumholz, M.~R., Ostriker, E.~C., et al.\ 2012, \mnras, 421, 3127.

    \bibitem[Novak et al.(2011)]{Novak11} Novak, G.~S., Ostriker, J.~P., \& Ciotti, L.\ 2011, \apj, 737, 26.

    \bibitem[Olivares et al.(2022)]{Olivares22} Olivares, V., Su, Y., Nulsen, P., et al.\ 2022, \mnras, 516, L101.

    \bibitem[Orienti \& Prieto(2010)]{Orienti10} Orienti, M. \& Prieto, M.~A.\ 2010, \mnras, 401, 2599.

    \bibitem[Pence \& Blackman(1984)]{Pence84} Pence, W.~D. \& Blackman, C.~P.\ 1984, \mnras, 210, 547.

    \bibitem[Pereira-Santaella et al.(2024)]{Pereira-Santaella24} Pereira-Santaella, M., Gonz{\'a}lez-Alfonso, E., Garc{\'\i}a-Bernete, I., et al.\ 2024, \aap, 681, A117.

    \bibitem[Peralta de Arriba et al.(2023)]{PeraltadeArriba23} Peralta de Arriba, L., Alonso-Herrero, A., Garc{\'\i}a-Burillo, S., et al.\ 2023, \aap, 675, A58.

    \bibitem[Pereira-Santaella et al.(2013)]{Pereira-Santaella13} Pereira-Santaella, M., Spinoglio, L., Busquet, G., et al.\ 2013, \apj, 768, 55.

    \bibitem[Pereira-Santaella et al.(2014)]{Pereira-Santaella14} Pereira-Santaella, M., Spinoglio, L., van der Werf, P.~P., et al.\ 2014, \aap, 566, A49.

    \bibitem[Pereira-Santaella et al.(2018)]{Pereira-Santaella18} Pereira-Santaella, M., Colina, L., Garc{\'\i}a-Burillo, S., et al.\ 2018, \aap, 616, A171.

    \bibitem[Pereira-Santaella et al.(2022)]{Pereira-Santaella22} Pereira-Santaella, M., {\'A}lvarez-M{\'a}rquez, J., Garc{\'\i}a-Bernete, I., et al.\ 2022, \aap, 665, L11.

    \bibitem[Prieto et al.(2014)]{Prieto14} Prieto, M.~A., Mezcua, M., Fern{\'a}ndez-Ontiveros, J.~A., et al.\ 2014, \mnras, 442, 2145.

    \bibitem[Randriamampandry et al.(2015)]{Randriamampandry15} Randriamampandry, T.~H., Combes, F., Carignan, C., et al.\ 2015, \mnras, 454, 3743.

    \bibitem[Ramos Almeida et al.(2009)]{RamosAlmeida09} Ramos Almeida, C., P{\'e}rez Garc{\'\i}a, A.~M., \& Acosta-Pulido, J.~A.\ 2009, \apj, 694, 1379.

    \bibitem[Ramos Almeida et al.(2022)]{Ramos Almeida22} Ramos Almeida, C., Bischetti, M., Garc{\'\i}a-Burillo, S., et al.\ 2022, \aap, 658, A155.

    \bibitem[Ramos Almeida et al.(2023)]{Ramos Almeida23} Ramos Almeida, C., Esparza-Arredondo, D., Gonz{\'a}lez-Mart{\'\i}n, O., et al.\ 2023, \aap, 669, L5.

    \bibitem[Reeves et al.(2007)]{Reeves07} Reeves, J.~N., Awaki, H., Dewangan, G.~C., et al.\ 2007, \pasj, 59, 301.

    \bibitem[Regan \& Teuben(2003)]{Regan03} Regan, M.~W. \& Teuben, P.\ 2003, \apj, 582, 723.

    \bibitem[Rieke et al.(2015)]{Rieke15} Rieke, G.~H., Ressler, M.~E., Morrison, J.~E., et al.\ 2015, \pasp, 127, 665.

    \bibitem[Rigopoulou et al.(2002)]{Rigopoulou02} Rigopoulou, D., Kunze, D., Lutz, D., et al.\ 2002, \aap, 389, 374.

    \bibitem[Rodr{\'\i}guez-Ardila et al.(2005)]{RodriguezArdila05} Rodr{\'\i}guez-Ardila, A., Riffel, R., \& Pastoriza, M.~G.\ 2005, \mnras, 364, 1041.

    \bibitem[Roussel et al.(2007)]{Roussel07} Roussel, H., Helou, G., Hollenbach, D.~J., et al.\ 2007, \apj, 669, 959.

    \bibitem[Runco et al.(2020)]{Runco20} Runco, J.~N., Malkan, M.~A., Fern{\'a}ndez-Ontiveros, J.~A., et al.\ 2020, \apj, 905, 57.

    \bibitem[Ruschel-Dutra et al.(2021)]{Ruschel-Dutra21} Ruschel-Dutra, D., Storchi-Bergmann, T., Schnorr-M{\"u}ller, A., et al.\ 2021, \mnras, 507, 74.

    \bibitem[Sanders \& Tubbs(1980)]{Sanders80} Sanders, R.~H. \& Tubbs, A.~D.\ 1980, \apj, 235, 803.

    \bibitem[Schmidt et al.(2019)]{Schmidt19} Schmidt, E.~O., Mast, D., D{\'\i}az, R.~J., et al.\ 2019, \aj, 158, 60.

    \bibitem[Shimizu et al.(2019)]{Shimizu19} Shimizu, T.~T., Davies, R.~I., Lutz, D., et al.\ 2019, \mnras, 490, 5860.

    \bibitem[Shipley et al.(2016)]{Shipley16} Shipley, H.~V., Papovich, C., Rieke, G.~H., et al.\ 2016, \apj, 818, 60.

    \bibitem[Solomon \& Vanden Bout (2005)]{Solomon05} Solomon, P. M., \& Vanden Bout, P. A. 2005, ARA\&A, 43, 677.

    \bibitem[Sormani et al.(2024)]{Sormani24} Sormani, M.~C., Sobacchi, E., \& Sanders, J.~L.\ 2024, \mnras, 528, 5742.

    \bibitem[Speranza et al.(2024)]{Speranza24} Speranza, G., Ramos Almeida, C., Acosta-Pulido, J.~A., et al.\ 2024, \aap, 681, A63.

    \bibitem[Stone et al.(2016)]{Stone16} Stone, M., Veilleux, S., Mel{\'e}ndez, M., et al.\ 2016, \apj, 826, 111.

    \bibitem[Sturm et al.(2011)]{Sturm11} Sturm, E., Gonz{\'a}lez-Alfonso, E., Veilleux, S., et al.\ 2011, \apjl, 733, L16.

    \bibitem[Tremaine et al.(2002)]{Tremaine02} Tremaine, S., Gebhardt, K., Bender, R., et al.\ 2002, \apj, 574, 740.

    \bibitem[V{\'e}ron-Cetty \& V{\'e}ron(2006)]{Veron-Cetty06} V{\'e}ron-Cetty, M.-P. \& V{\'e}ron, P.\ 2006, \aap, 455, 773.

    \bibitem[Wright et al.(2015)]{Wright15} Wright, G.~S., Wright, D., Goodson, G.~B., et al.\ 2015, \pasp, 127, 595.

    \bibitem[Wright et al.(2023)]{Wright23} Wright, G.~S., Rieke, G.~H., Glasse, A., et al.\ 2023, \pasp, 135, 048003.

    \bibitem[Zhang \& Ho(2023)]{Zhang23} Zhang, L. \& Ho, L.~C.\ 2023, \apjl, 953, L9.

    \bibitem[Zhang et al.(2024a)]{Zhang24a} Zhang, L., Packham, C., Hicks, E.~K.~S., et al.\ 2024, \apj, 974, 195.
    
    \bibitem[Zhang et al.(2024)]{Zhang24b} Zhang, L., Garc{\'\i}a-Bernete, I., Packham, C., et al.\ 2024, \apjl, 975, L2.

    \bibitem[Zoghbi et al.(2014)]{Zoghbi14} Zoghbi, A., Cackett, E.~M., Reynolds, C., et al.\ 2014, \apj, 789, 56.

\end{thebibliography}
%

\begin{appendix}
\onecolumn
\section{Comparison between ALMA and HST maps.}
\label{app:ALMAvsHST}
\begin{figure}[ht]
\centering
\includegraphics[width=1\columnwidth, trim={0 0 0 0},clip]{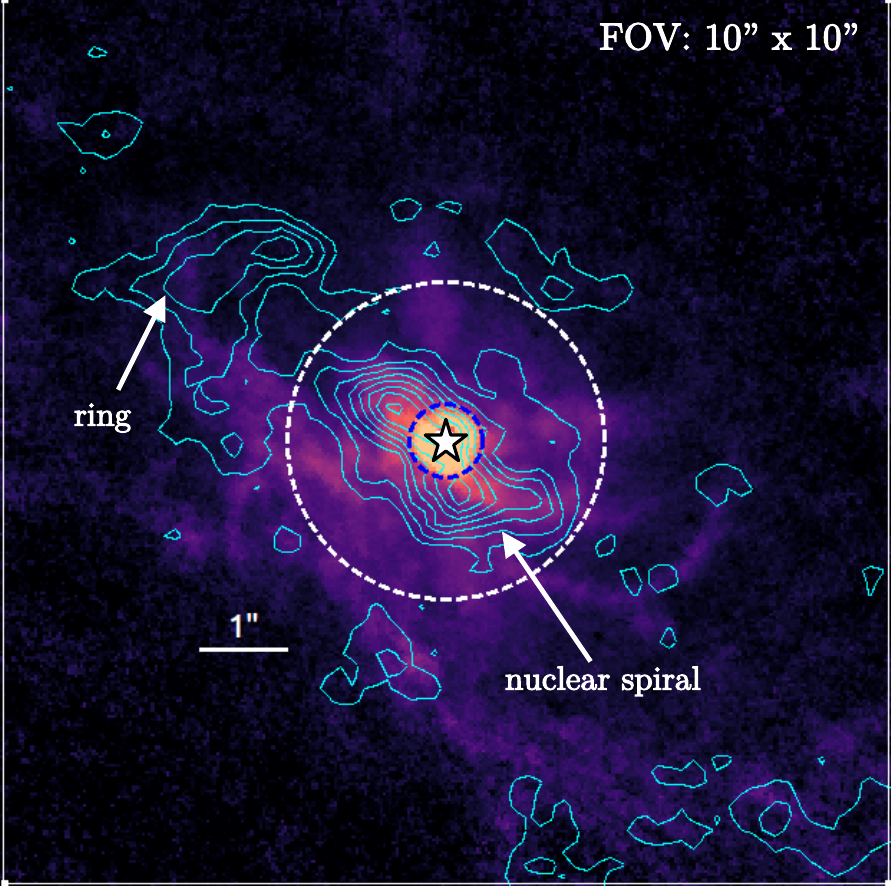} 
\caption{V-H color map obtained using HST/WFC3 images in the F606W and F160W filters. The cyan solid contours correspond to 2.2$\times 10^{-3}$, 4.4$\times 10^{-3}$, 6.6$\times10^{-3}$, 8.8$\times10^{-3}$, 1.1$\times 10^{-2}$, 1.3$\times 10^{-2}$, 1.5$\times 10^{-2}$, 1.7$\times 10^{-2}$, and 2$\times 10^{-2}$ [Jy km/s] ALMA CO(2-1) intensity. The black and blue dashed circles correspond to the regions used to extract the nuclear (diameter of 0\farcs8 - 2\farcs5 $\simeq$ 141-440 pc) and extended (diameter of 3\farcs6 $\simeq$ 634 pc) JWST/MIRI spectra shown in Fig. \ref{figure:MIRISpectra}.}
\label{Fig:ALMA_vs_HST}
\end{figure}
\clearpage

\section{Moment maps of CO(2$-$1), $\rm{H_2}$0$-$0S(1), and $\rm{H_2}$0$-$0S(3).}
\label{app:moments}

\begin{figure}[ht]
\centering
\includegraphics[width=1\columnwidth, trim={5 2 15 5},clip]{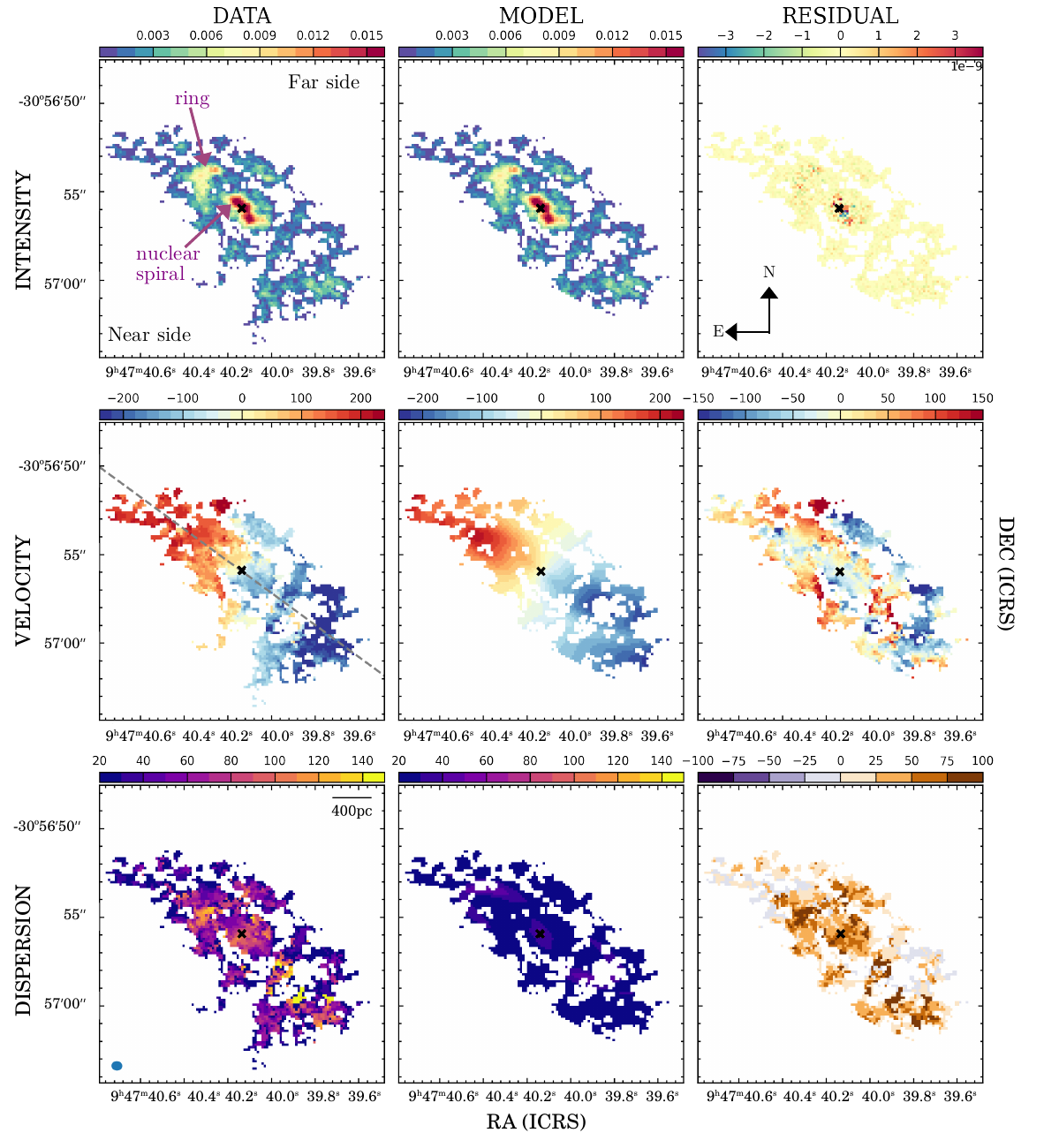}
\caption{From left to right: CO(2$-$1) observed moment maps, BAROLO model and residuals. From top to bottom moment 0 [Jy km/s], 1 [km/s], and 2 [km/s]. The kinematic major axis (PA=56$^{\circ}$) is shown with a dashed black line, and the beam size (0\farcs39) is indicated in the bottom left corner of the observed moment 2 map.}
\label{Fig:MomentMapCO}
\end{figure}

\begin{figure}[ht]
\centering
\includegraphics[width=1\columnwidth, trim={0 2 20 2},clip]{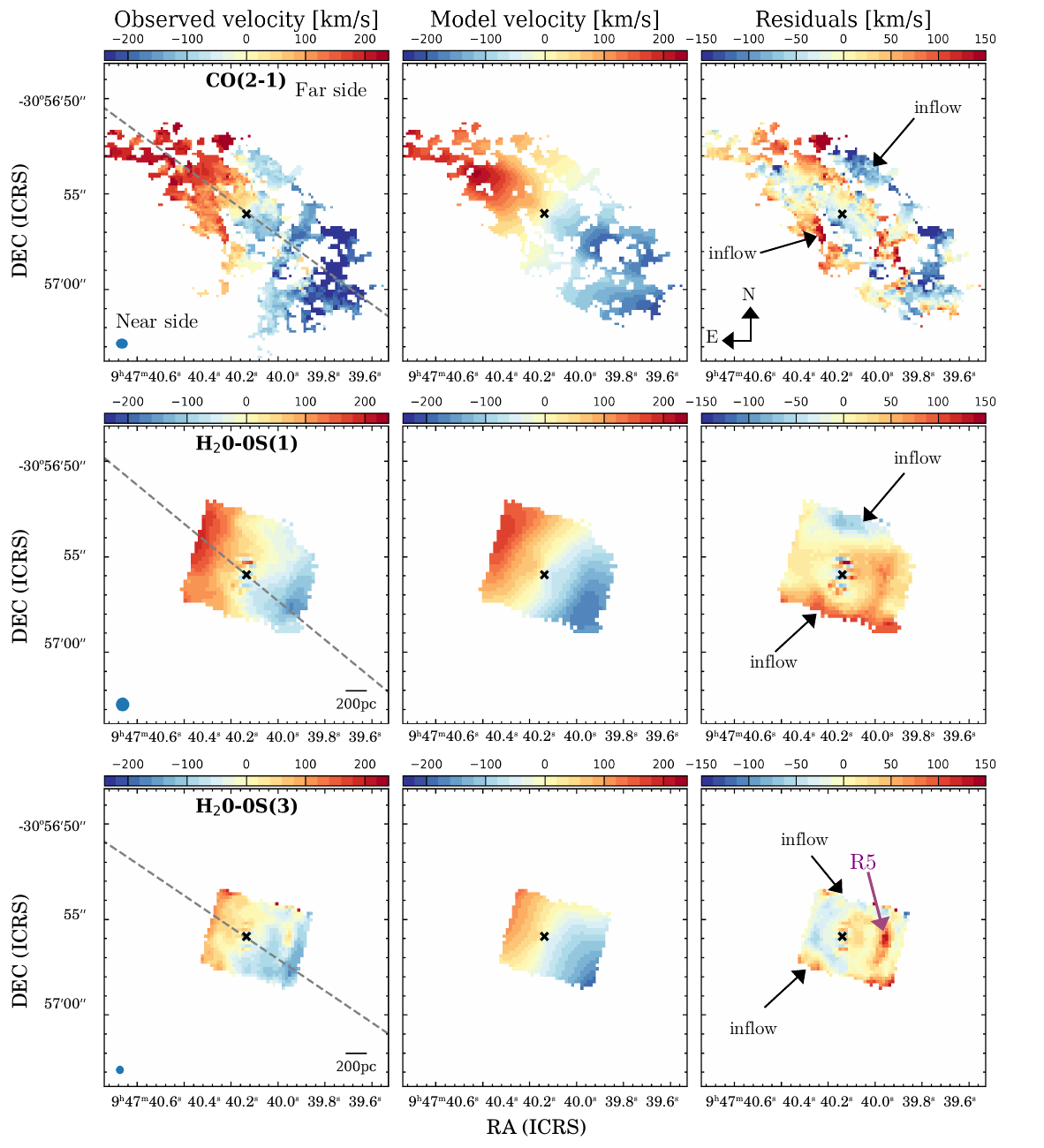}
\caption{From top to botton, moment 1 maps of the CO(2$-$1), H$_2$0$-$0S(1), and H$_2$0$-$0S(3) emission lines. From left to right: data, \barolo~model, and residuals from subtracting the model from the data. The kinematic major axis determined by \barolo~is indicated as a dashed gray line, and the corresponding beam sizes are shown as blue ellipses in the bottom left corner of the left panels. The black cross corresponds to the peak of the ALMA continuum.}
\label{Fig:MomentMapvel}
\end{figure}

\begin{figure}[ht]
\centering
\includegraphics[width=1.\columnwidth, trim={5 2 0 2},clip]{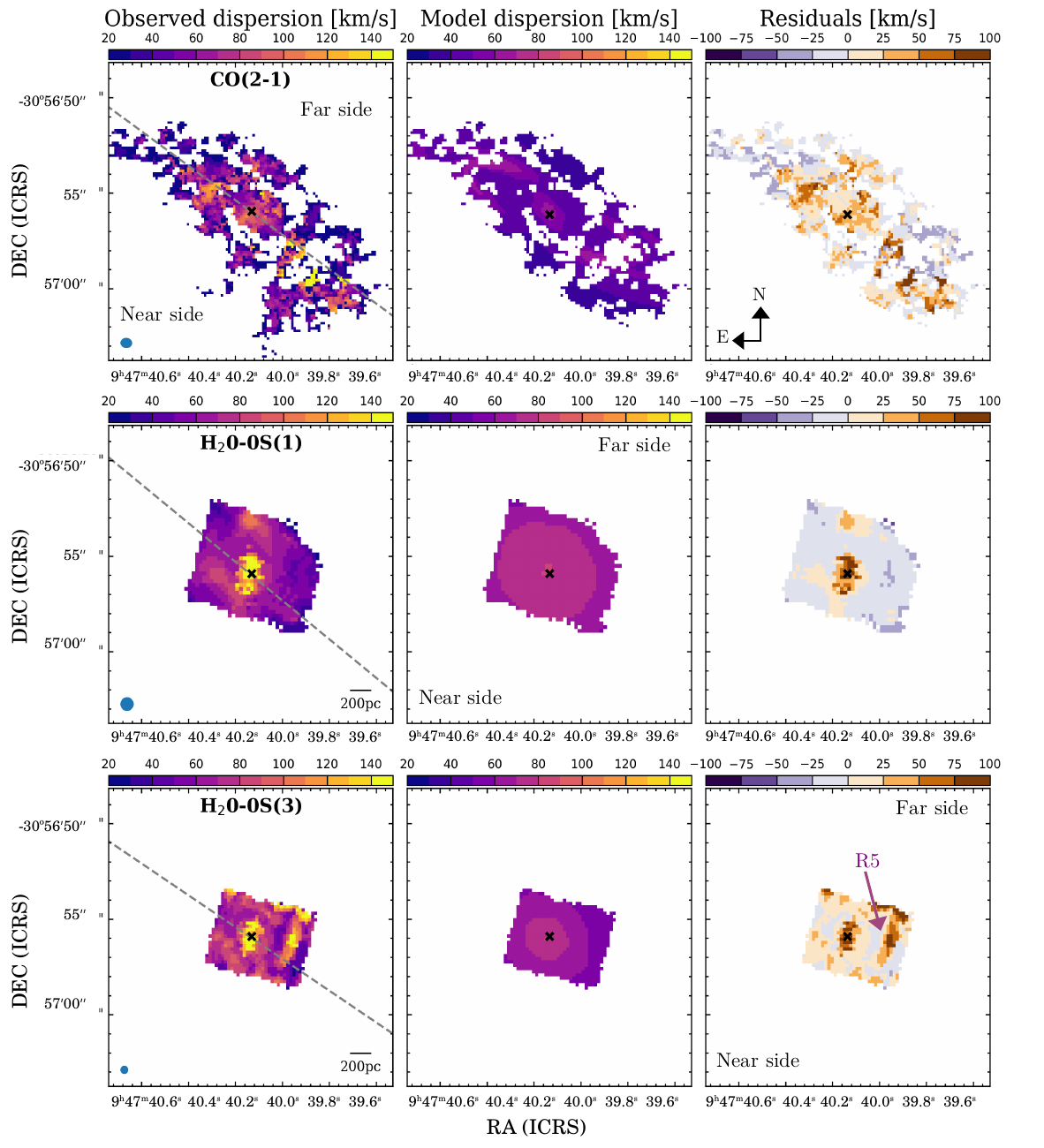} 
\caption{Same as in Fig. \ref{Fig:MomentMapvel}, but for the moment 2 maps.}
\label{Fig:MomentMapdisp}
\end{figure}

\end{appendix}

\end{document}